\documentclass[fleqn,usenatbib]{mnras}

\usepackage{newtxtext,newtxmath}
\usepackage[T1]{fontenc}

\DeclareRobustCommand{\VAN}[3]{#2}
\let\VANthebibliography\thebibliography
\def\thebibliography{\DeclareRobustCommand{\VAN}[3]{##3}\VANthebibliography}

\usepackage{graphicx}	% Including figure files
\usepackage{amsmath}	% Advanced maths commands
\usepackage{xspace}
\usepackage{gensymb}
\usepackage{pdflscape}
\usepackage{xcolor}

\newcommand{\oc}{$\omega$\ Cen\xspace}  % Omega Centauri
\newcommand{\ergs}{erg~s$^{-1}$\xspace}
\newcommand{\chandra}{{\it Chandra}\xspace}
\newcommand{\msun}{M$_{\sun}$\xspace}

\title[X-ray MSPs in \oc]{A \chandra X-ray study of millisecond pulsars in the globular cluster Omega Centauri: a correlation between spider pulsar companion mass and X-ray luminosity}

\author[J. Zhao \& C. O. Heinke]{
Jiaqi Zhao,$^{1}$\thanks{E-mail: jzhao11@ualberta.ca}
Craig O. Heinke$^{1}$
\\
$^{1}$ Physics Dept., CCIS 4-183, University of Alberta, Edmonton, AB, T6G 2E1, Canada\\
}

\date{Accepted XXX. Received YYY; in original form ZZZ}

\pubyear{2023}

\begin{document}
\label{firstpage}
\pagerange{\pageref{firstpage}--\pageref{lastpage}}
\maketitle

\begin{abstract}

Millisecond pulsars (MSPs) are faint X-ray sources commonly observed in Galactic globular clusters (GCs). 
In this work, we 
%utilize the latest radio detections and timing solutions of
investigate 18 MSPs newly found in the GC Omega Centauri (\oc) and search for their X-ray counterparts using \chandra observations with a total exposure time of 290.9 ks. 
We identify confident X-ray counterparts for 11 of the MSPs, with 9 of them newly identified in this work based on their positions,
spectral properties, and X-ray colours. 
The X-ray spectra of 9 MSPs are well described by a neutron star hydrogen atmosphere model, while 2 MSPs are well fitted by a power-law model. 
The identified MSPs have X-ray luminosities ranging from $1.0\times10^{30}$ \ergs to $1.4\times10^{31}$ \ergs. 
Additionally,
for population comparison purposes,
we study the X-ray counterpart to MSP E in the GC M71, and find its X-ray spectrum is well described by blackbody-like models with a luminosity of $1.9\times10^{30}$ \ergs.
We investigate the empirical correlations between X-ray luminosities and minimum companion masses, as well as mass functions, of spider pulsars. 
Clear correlations are observed, with best-fit functions of 
$\log_{10}{L_X} = (1.0\pm0.1) \log_{10}{M_{c, min}} + (32.5\pm0.2)$ and $\log_{10}{L_X} = (0.35\pm0.04) \log_{10}{\rm MF} + (32.71\pm0.20)$, respectively, with an intrinsic scatter of $\log_{10}{L_X}$ of $\sim$0.3, where $L_X$ is the 0.5--10 keV X-ray luminosity, $M_{c, min}$ is the minimum companion mass, and MF represents the mass function, in solar masses.
% \jakeSecond{(Current word counts: $\sim$243)}

\end{abstract}


%%%%%%%%%%%%%%%%%%%%%%%%%%%%%%%%%%%%%%%%%%%%%%%%%%

%%%%%%%%%%%%%%%%% BODY OF PAPER %%%%%%%%%%%%%%%%%%

\section{Introduction}
\label{sec:intro}

Given the high stellar densities in the core regions of globular clusters (GCs), stellar interactions in GCs are highly significant, such as tidal captures \citep{Fabian1975} and stellar exchanges \citep{Hills1976}. 
Consequently, 
low-mass X-ray binaries (LMXBs) 
are overabundantly produced in GCs \citep[e.g.,][]{Katz75, Verbunt87,Pooley06,Heinke2006Ter5}, and GCs with higher stellar encounter rates are expected to host larger populations of such systems \citep{Verbunt87,Bahramian2013}.
LMXBs are the progenitors of millisecond pulsars (MSPs), where the neutron star (NS) accretes mass and angular momentum from its companion star and hence spins up to a period of a few milliseconds \citep{Alpar1982,Papitto2013}. 
GCs are therefore the ideal birthplaces for MSPs as well. 
Indeed, 280 pulsars have been discovered in 38 Galactic GCs to date,  262 of which are MSPs (i.e. spin periods $\lesssim$ 30 ms), with new identifications occurring frequently \citep[e.g.][]{Lian23,Pan2023,Chen2023}.\footnote{See Paulo Freire's GC pulsar catalogue at \url{https://www3.mpifr-bonn.mpg.de/staff/pfreire/GCpsr.html} for up-to-date information.}

MSPs %in GCs 
are commonly observed 
%with companion stars
to be hosted in binary systems, which is not surprising considering their evolution.
%Specifically, 
MSPs coupled with non-/semi-degenerate low-mass stars are the so-called spider pulsars. 
%It has been well established that 
The companion masses of spider pulsars have a bi-modal distribution, which divides them into two groups: redbacks (RBs) with companion masses between $\sim$0.1 \msun and $\sim$0.5 \msun, and black widows (BWs) with companion masses less than $\sim$0.05 \msun\ \citep{Roberts2013}. 
%Furthermore, 
All of the 
%detected
known RBs show eclipses in their radio pulsations and around half of discovered BWs are radio eclipsing.
(Note that non-eclipsing RBs may exist, e.g. those with very low inclination angles, but might be erroneously classified as other binary types, such as MSPs associated with helium-core white dwarfs.)
%which 
These radio eclipses likely result from dispersion, scattering, and/or absorption in the stripped material of the companion \citep[see e.g.,][]{Stappers1996,Roy2015,Polzin2020}.
The canonical companions of MSPs %may also be
are white dwarfs (WDs), %compact objects, 
with helium-core WDs being the most commonly observed \citep[e.g.][]{Lorimer2008}. 
It is noteworthy that in GCs, the companions of MSPs might result from exchange interactions, rather than originating solely from the recycling phenomenon.
On the other hand, of order a third of MSPs are isolated, %in GCs, 
which are likely produced by merger of an MSP with its very low-mass companion due to dynamically unstable mass transfer \citep{Ivanova2008}. In globular clusters, MSPs can also become isolated through dynamical interactions \citep{VerbuntFreire14}.

MSPs are generally faint X-ray emitters, with typical X-ray luminosities ranging from $\sim1\times10^{30}$ \ergs to $1\times10^{32}$ \ergs, and up to $\sim1.4\times10^{33}$ \ergs in 0.3--8 keV energy band \citep[see][]{Zhao2022}. 
Based on the likely origins, X-ray emission from MSPs can be categorised into three groups. 
Thermal X-ray emission is considered to be originated from the hotspots on NS surface (presumably near the NS magnetic poles), where are heated up by returning flows of relativistic particles accelerated in magnetosphere \citep{Harding2002}. 
Thermal emission is commonly observed from isolated MSPs and MSPs coupled with white dwarfs, with blackbody-like spectra \citep[e.g.][]{Zavlin02,Bogdanov2006}.
%Recent 
X-ray observations have also shown broad pulsations and modulations from thermally emitting MSPs \citep[e.g.][]{Zavlin98,Guillot2019,Wolff2021}, providing evidence for the hotspot origin. 
X-rays generated from pulsar magnetosphere are believed to result in the observed narrow X-ray pulsations from only a few very energetic MSPs, such as PSR M28 A \citep[see e.g.,][]{Saito1997,Bogdanov2011}, PSR J0218+4232 \citep{Kuiper98}, and PSR B1937+21 \citep{Takahashi01}. 
This type of X-ray emission is predominantly observed with a hard power-law spectrum, indicating a non-thermal origin. 
The third type of X-ray emission from MSPs is typically seen from spider pulsars, likely produced by intrabinary shocks, where the relativistic pulsar winds collide with the outflow material from the companion star \citep[see e.g.,][]{Arons1993,Wadiasingh2017,Kandel2019}. 
In addition, it is found that RBs are X-ray luminous in the MSP population, and eclipsing BWs generally brighter than non-eclipsing BWs in X-rays \citep{Zhao2022}.

Omega Centauri (NGC 5139; hereafter \oc) is the most massive %GC
stellar system in the Galaxy found so far, with a stellar mass of $3.6\times10^6$ \msun \citep{Baumgardt2018}. 
It is located in the constellation of Centaurus at a distance of 5.43 kpc from the Sun \citep{Baumgardt2021}.\footnote{See also Baumgardt's catalogue at \url{https://people.smp.uq.edu.au/HolgerBaumgardt/globular/} for other fundamental parameters of \oc.}
Given its outstanding stellar population, observations in different bands towards \oc are of great interest. 
For example, \citet{Abdo2010} analysed $\gamma$-ray observations from {\it Fermi} Large Area Telescope and found that \oc exhibits MSP spectral properties in $\gamma$-rays, even though no MSP had been discovered therein at that time. 
Nonetheless, they predicted a population of 19$\pm$9 MSPs to explain the observed $\gamma$-rays. 
\citet{Henleywillis2018} conducted a deep X-ray survey of \oc using \chandra observations and identified a total of 233 X-ray sources, with 45 likely optical counterparts. 
In addition, they found more than 30 unidentified X-ray sources that are plausible MSP candidates given the similar X-ray properties to other MSPs in GCs. 
Although the presence of MSPs in \oc had been predicted and searched for over decades \citep[e.g.][]{Edwards2001,Possenti2005}, it was only in recent years that MSPs were discovered in this cluster, suggesting a lack of sensitivity of previous radio detections. 
The first five MSPs, including four isolated MSPs and one eclipsing BW, were detected by \citet{Dai2020} using Parkes radio telescope (also known as ``Murriyang''). 
More recently, \citet{Chen2023} discovered an additional 13 MSPs in \oc using MeerKAT radio telescope, bringing the total number of MSPs discovered in this cluster to 18 
(including 10 isolated MSPs, five BWs with three of them showing eclipses, and three binary MSPs with undetermined companion nature),
which makes \oc the Galactic GC with the third largest detected MSP population.
During the preparation of this work, \citet{Dai2023} published new timing solutions of the first five discovered MSPs \citep{Dai2020}. 

Motivated by recent radio detections and timing solutions of MSPs in \oc, in this paper, we conduct a search for their X-ray counterparts and analyse the corresponding X-ray spectra using \chandra observations. 
Additionally, we empirically investigate the correlations between X-ray luminosities and  companion masses, as well as between X-ray luminosities and the mass functions, for spider MSPs.
This is best done with globular cluster pulsars, since only for these pulsars do we generally have accurate distances and thus X-ray luminosities. 
This paper is organised as follows. 
In Section~\ref{sec:obseration}, we describe the  \chandra observations used, and the process of data reduction. 
In Section~\ref{sec:results}, we compare our X-ray detections with existing catalogues, analyse the X-ray spectra and properties, and identify X-ray counterparts to MSPs.
We discuss factors that may influence X-ray luminosities of spider pulsars in Section~\ref{sec:discussion}, and draw conclusion in Section~\ref{sec:conclusions}.

\section{Observations and data reduction}
\label{sec:obseration}

We obtained four \chandra X-ray observations of \oc from the \chandra Data Archive\footnote{\url{https://cxc.cfa.harvard.edu/cda/}}, with a total exposure time of 290.9 ks (see Table~\ref{tab:obs_oc}). All observations were taken using the Advanced CCD Imaging Spectrometer (ACIS) with chip numbers from 0 to 3, i.e. ACIS-I chips, and data were recorded and configured in VFAINT mode. 

We performed data reduction and analysis using {\sc ciao}\footnote{Chandra Interactive Analysis of Observations, publicly available at \url{https://cxc.harvard.edu/ciao/}} \citep[version 4.15.1 with {\sc caldb} 4.10.2;][]{Fruscione2006}. For each observation, we first reprocessed the dataset using the {\tt chandra\_repro} script to create a new level=2 event file and a new bad pixel file. We then corrected the absolute astrometry by cross-matching the second brightest X-ray source in \oc \citep[source ID 94a, assigned in][]{Haggard2009} with its optical counterpart \citep[see][and references therein]{Henleywillis2018}. Specifically, we ran {\tt wavdetect}, a Mexican-Hat Wavelet source detection algorithm in {\sc ciao}, to detect and extract the X-ray centroid of 94a in each observation. The X-ray position was used to match the position of its optical counterpart at R.A.=13:16:01.61, Dec.=$-$47:33:05.7 \citep{vanLeeuwen2000} using {\tt wcs\_match}, and then the obtained offset information was applied to correct and update the absolute astrometry for each dataset using {\tt wcs\_update} script. The average offset for the four observations was around 0.43 arcsec.

\begin{figure*}
    \centering
    \includegraphics[width=0.93\textwidth]{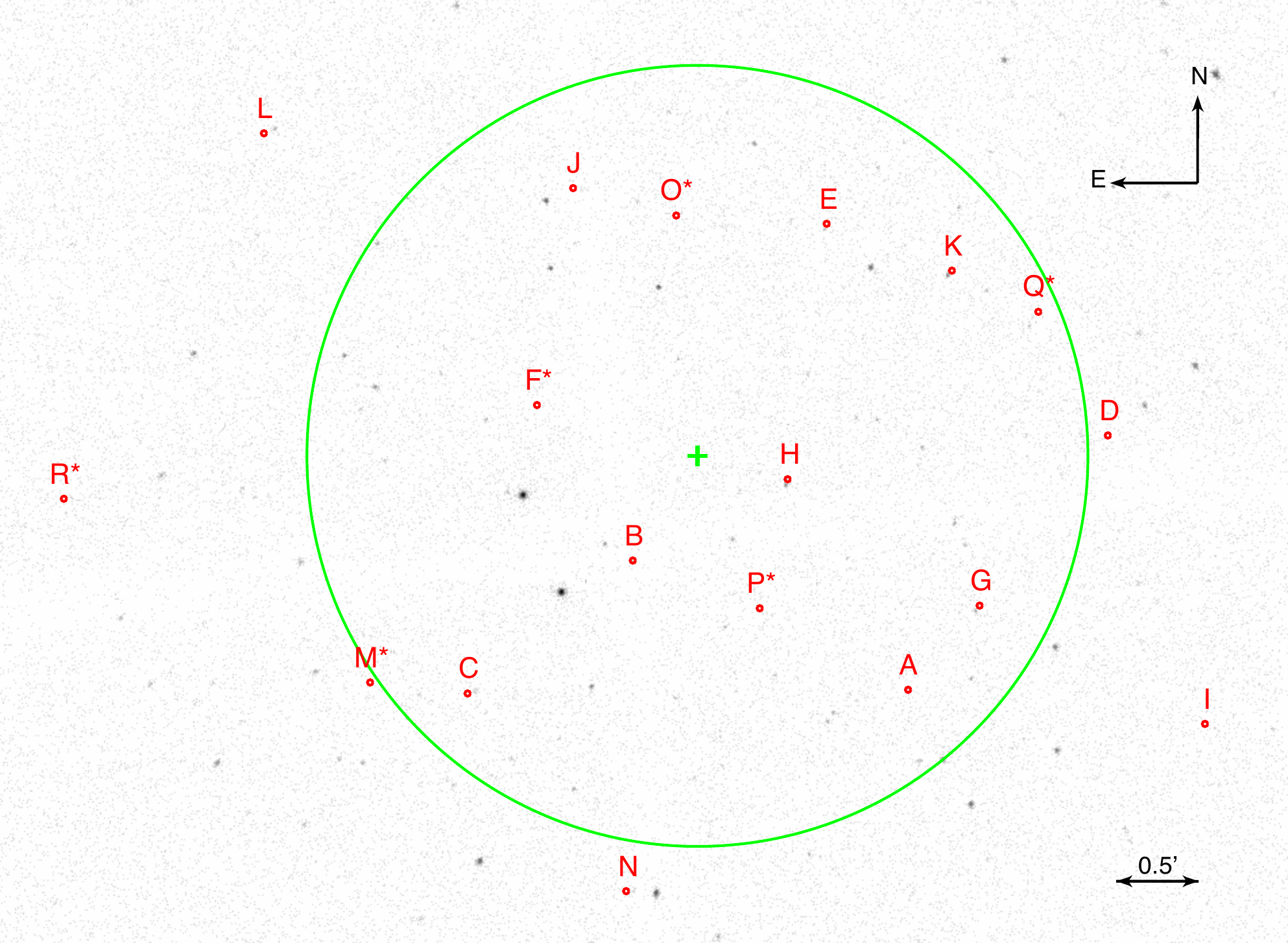}
    \includegraphics[width=0.95\textwidth]{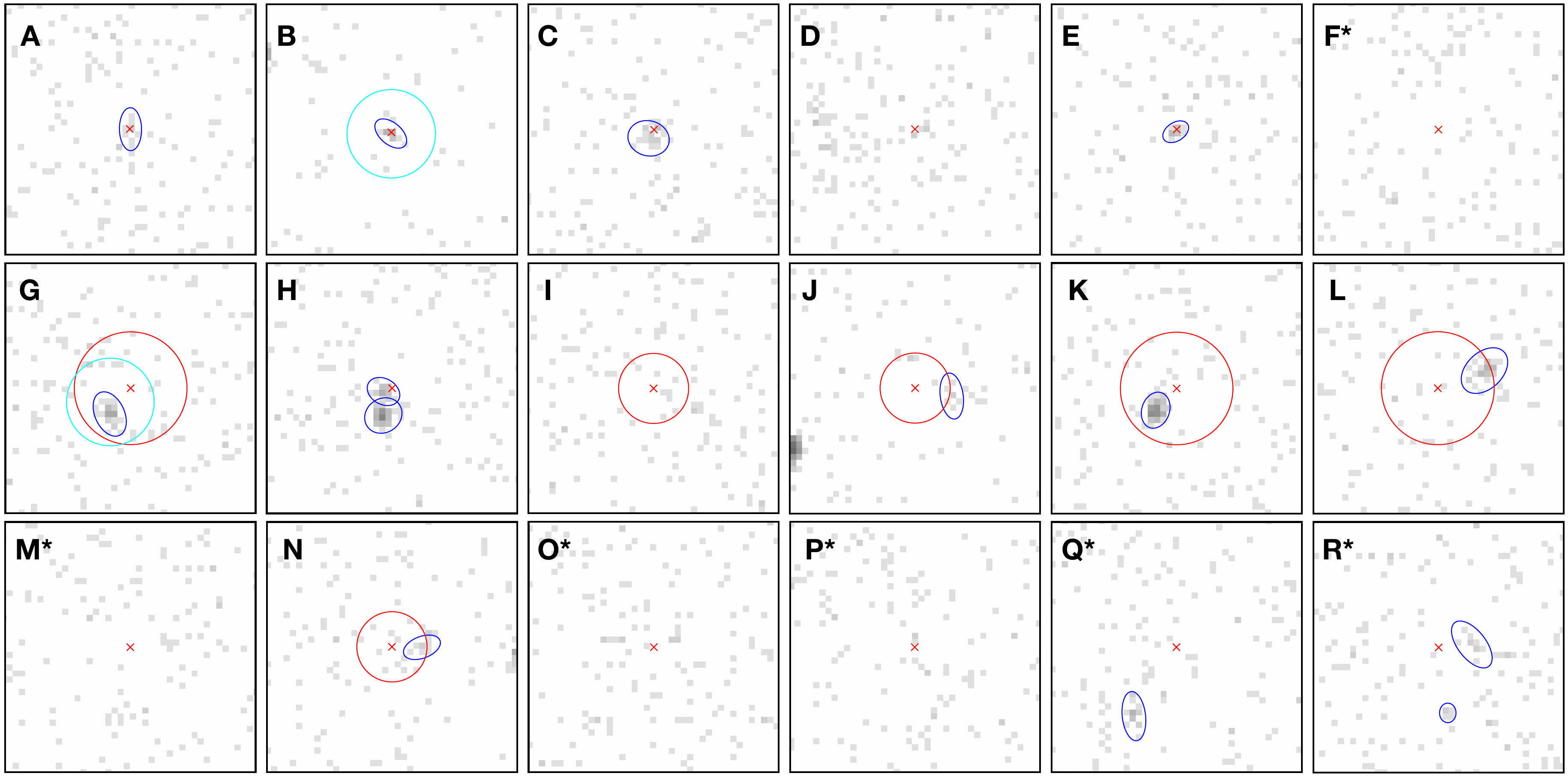}
    \caption{{\it Upper panel}: merged \chandra X-ray image of the core region of \oc in the band 0.3--8 keV. Red circles (1-arcsec radii) indicate radio positions of the 18 discovered MSPs from \citet{Chen2023} and \citet{Dai2023}, with MSP names labelled. Green cross and larger circle show the centre of \oc at R.A.=13:26:47.24, Dec.=$-$47:28:46.5, and the 2.37-arcmin core radius, respectively \citep[2010 edition]{Harris1996}. {\it Lower panels}: zoom-in X-ray images centred at the radio positions of MSPs (labelled with red crosses), shown in 20$\times$20 arcsec$^2$ boxes. Blue ellipses show the candidate X-ray counterparts to the corresponding MSPs detected by {\tt wavdetect}. For MSPs G, I, J, K, L, and N, the red circles indicate corresponding search area for X-ray counterparts, while for MSPs F, H, M, O, P, Q, and R, the search regions are the whole box area, respectively (see Section~\ref{sec:obseration} for details). No search circles are shown for MSPs A--E given their precise timing positions. For MSPs B and G, their corresponding ATCA counterparts are shown with cyan circles with a radius of 3.5 arcsec \citep[see][]{Dai2020}. MSP names labelled with asterisks denote their radio positions were placed at the centres of the beams \citep[see][]{Chen2023}. North is up, and east is to the left. } 
    \label{fig:x-ray_img}
\end{figure*}

To find the X-ray counterparts to the MSPs in \oc, we first merged the four \chandra observations using {\tt merge\_obs} script, and applied {\tt wavdetect} again in the merged 0.3--8 keV X-ray image (see the upper panel in Figure~\ref{fig:x-ray_img}), with a scale list of [1, 1.4, 2, 4, 8], and a significance threshold of 10$^{-5}$. The parameter settings used in this case were designed to achieve detection completeness and increase the ability to detect weaker sources (though spurious detections potentially increased too). 
Using the radio positions of the MSPs presented in \citet[for MSPs A--E]{Dai2023} and \citet[for the other 13 MSPs]{Chen2023}, we searched for X-ray sources detected by {\tt wavdetect} in the vicinities of the discovered MSPs.

%Specifically, given the precise 
The timing positions of MSPs A--E \citep{Dai2023} are of such high precision ($\ll$0.1") that the X-ray position errors (typically $<$1") dominate. 
Indeed, the positions of four detected counterparts to MSPs A, B, C, and E, respectively, are all consistent with the timing positions of corresponding MSPs, within their 95\% X-ray positional uncertainties (see Table~\ref{tab:properties}). It also validates our astrometric correction, though an average offset of 0.3 arcsec between radio and X-ray positions of these four MSPs is still present.
Hence, we suggest these four X-ray sources are genuine counterparts to MSPs A, B, C, and E, respectively.
MSP D, however, has no detected X-ray counterpart, likely due to its intrinsically faint nature.

For the other 13 MSPs detected by MeerKAT \citep{Chen2023}, we considered them in two groups: group one contains MSPs having multibeam detections and hence relatively good localisations using SeeKAT, including MSPs G, I, J, K, L, and N; group two denotes those MSPs 
detected in only one MeerKAT beam,
including MSPs F, H, M, O, P, Q, and R. 
We note that, despite MSP H having multibeam detections, its localisation was poorly constrained with uncertainties similar to other group two MSPs. Therefore, we considered it a member of the group two MSPs.
The typical positional uncertainties (2-$\sigma$) of MSPs in group one are 1.5 arcsec in R.A. and 1 arcsec in Dec., respectively, while for group two MSPs, the typical positional uncertainties are 15 arcsec in R.A and 6 arcsec in Dec., respectively \citep{Chen2023}.
We also compared the difference between the precise timing positions from \citet{Dai2023} and the positions localised with SeeKAT from \citet[see their Table 2]{Chen2023} for MSPs A--E, and found average offsets of 2.2 arcsec in R.A. and 1.2 arcsec in Dec., respectively. 
The largest separation between a timing position and a SeeKAT position was 3.1 arcsec for MSP C, about 1.4 times the quoted SeeKAT uncertainty, so we conservatively adopted two times the quoted SeeKAT uncertainties to define search radii for MSPs in group one.
Considering additionally 
the X-ray positional uncertainties (see Table~\ref{tab:properties}),
we found one counterpart to each of MSPs G, J, K, L and N, 
%and two counterparts to MSP H, 
whereas no counterpart was detected to MSP I within the region (see the lower panels in Figure~\ref{fig:x-ray_img}).

It is more difficult to identify the X-ray counterparts to MSPs in group two, given their large positional uncertainties. 
Nonetheless, since these MSPs were positioned at the centres of the beams with the brightest detections, while the synthesised beam was shaped as an approximate ellipse with major and minor axis of 20.5 arcsec and 9.3 arcsec, respectively \citep{Chen2023}, we searched for X-ray counterparts for MSPs in group two within 20\arcsec$\times$20\arcsec\ regions. 
We identified one potential X-ray counterparts to MSP Q, and two potential X-ray counterparts for each of MSPs H and R, while we found no potential counterparts for MSPs F, M, O, or P 
(see the lower panels in Figure~\ref{fig:x-ray_img}).  

To further analyse the X-ray properties and identify the nature of these potential X-ray counterparts, we extracted their spectra from 1-arcsec radii circles centred at the centroids obtained by {\tt wacdetect} for each of the four observations individually, using the {\tt specextract} script. 
We note that, since \chandra observations of \oc were operated in two different cycles, which were separated by $\sim$12 years, the observed spectra in $<$2~keV band may be significantly different from each of the obtained ones in the two cycles, due to the molecular contamination on the ACIS filters\footnote{See \url{https://cxc.harvard.edu/proposer/POG/html/chap6.html} for more details.}
and/or the intrinsic long-term variability in X-rays \citep[e.g. PSR J1748$-$2446A in Terzan 5; see][]{Bogdanov2021}.
Thus, instead of combining all of the spectra extracted from those four observations for each X-ray source, we combined their X-ray spectra separately according to the observing cycles, i.e. cycles 1 and 13.
We then fitted the two combined spectra in two cycles simultaneously for each X-ray source with parameters forced to be the same. 
We note that, however, the spectral extractions of a1 and e1 were not successful in Observation 653, while the extraction of n1 was unsuccessful in Observation 13727, due to zero counts in the extraction region, which likely reflects the varied exposure times (see Table~\ref{tab:obs_oc}).
For these three sources, we manually added the corresponding exposure times in their spectral information files before spectral analysis. 
%(68.6~ks for cycle 1 observations, and 222.2 ks for cycle 12 observations).  
In particular, the two X-ray sources detected around MSP R, r1 and r2, only had extracted spectra from the two cycle 13 observations. Their spectral properties therefore only reflect latter observations.
Background spectra were also extracted from source-free regions for each MSP.
We also note that \citet{Zhao2022} have presented spectral analysis of MSPs A and B. Here we reanalysed their X-ray spectra to keep consistency in this work. 

\begin{table}
	\centering
	\caption{\chandra observations of \oc}
	\label{tab:obs_oc}
	\begin{tabular}{lcccr} % four columns, alignment for each
		\hline
	Observation	& & Date of & Observing & Exposure\\
		 ID & Instrument & Observation & Cycle & Time (ks) \\
		\hline
		653 & ACIS-I & 2000 Jan 24 & 01 & 25.0 \\
        1519 & ACIS-I & 2000 Jan 25 & 01 & 43.6 \\
        13726 & ACIS-I & 2012 Apr 17 & 13 & 173.7 \\
        13727 & ACIS-I & 2012 Apr 16 & 13 & 48.5 \\
		\hline
	\end{tabular}
\end{table}

\begin{table*}
    \centering
    \caption{Basic properties of the MSPs and X-ray counterparts in this work}
    \begin{tabular}{lllccccccc}
    \hline
MSP & \multicolumn{2}{c}{Radio Position$^a$} & X-ray$^b$ & \multicolumn{2}{c}{X-ray Position$^c$} & PU$^d$ & Offset$^e$ & Existing$^f$ \\%Extraction? $^f$ & 
Name	&	R.A. (hh:mm:ss)	&	Dec. (dd:mm:ss)	&	Name	&	R.A. (hh:mm:ss)	&	Dec. (dd:mm:ss)	& (arcsec) &	(arcsec)	&		ID \\
\hline
A	&	13:26:39.6699(2)	&	$-$47:30:11.641(3)	&	a1	&	13:26:39.66	&	$-$47:30:11.7	& 1.02 &	0.06	&	\\ %N/Y/Y/Y	&		\\
B	&	13:26:49.5688(3)	&	$-$47:29:24.889(4)	&	b1	&	13:26:49.57	&	$-$47:29:25.0	& 0.47 &	0.10	&	13d	\\ %Y/Y/Y/Y	&	
C	&	13:26:55.2219(6)	&	$-$47:30:11.753(9)	&	c1	&	13:26:55.26	&	$-$47:30:12.4	& 0.78 &	0.81	&	23g	\\ % Y/Y/Y/Y	&	
D	&	13:26:32.7130(2)	&	$-$47:28:40.053(3)	&	--	&	--	&	--	&	--	& -- &	\\ %--	&		\\
E	&	13:26:42.67844(7)	&	$-$47:27:23.999(1)	&	e1	&	13:26:42.69	&	$-$47:27:24.2	& 0.66 &	0.21	&	11f	\\ %N/Y/Y/Y	&	
F	&	13:26:53(1)	&	$-$47:28:28(6) *	&	--	&	--	&	--	&	--	& -- &	\\ %--	&		\\
G	&	13:26:37.1(2)	&	$-$47:29:41(1)	&	g1	&	13:26:37.27	&	$-$47:29:43.1	& 0.47 &	2.65	&	24f	\\ %Y/Y/Y/Y	&	
H	&	13:26:44(1)	&	$-$47:28:55(4)	&	h1	&	13:26:44.07	&	$-$47:28:55.3	& 0.40 &	0.73	&	14c	\\ %Y/Y/Y/Y	&	
	&		&		&	h2	&	13:26:44.07	&	$-$47:28:57.2	& 0.22 &	2.29	&	14c	\\ %Y/Y/Y/Y	&	
I	&	13:26:29.0(1)	&	$-$47:30:24(1)	&	--	&	--	&	--	&	--	& -- &	\\ %--	&		\\
J	&	13:26:51.7(1)	&	$-$47:27:09(1)	&	j1	&	13:26:51.41	&	$-$47:27:09.6	& 1.03 &	2.99	&	\\ %Y/Y/Y/Y	&		\\
K	&	13:26:38.1(1)	&	$-$47:27:39(2)	&	k1	&	13:26:38.27	&	$-$47:27:40.7	& 0.28 &	2.42	&	21d	\\ %Y/Y/Y/Y	&	
L	&	13:27:02.8(1)	&	$-$47:26:49(2)	&	l1	&	13:27:02.43	&	$-$47:26:47.6	& 0.69 &	3.98	&	32d	\\ %Y/Y/Y/Y	&	
M	&	13:26:59(1)	&	$-$47:30:09(6) *	&	--	&	--	&	--	&	--	& -- &	\\ %--	&		\\
N	&	13:26:49.8(1)	&	$-$47:31:25(1)	&	n1	&	13:26:49.56	&	$-$47:31:25.0	& 1.22 &	2.39	&	\\ %Y/Y/Y/N	&		\\
O	&	13:26:48(1)	&	$-$47:27:19(6) *	&	--	&	--	&	--	&	--	&	-- & \\ %--	&		\\
P	&	13:26:45(1)	&	$-$47:29:42(6) *	&	--	&	--	&	--	&	--	&	-- & \\ %--	&		\\
Q	&	13:26:35(1)	&	$-$47:27:54(6) *	&	q1	&	13:26:35.33	&	$-$47:27:59.5	& 0.71 &	6.47	&	\\ %Y/Y/Y/Y	&		\\
R	&	13:27:10(1)	&	$-$47:29:02(6) *	&	r1	&	13:27:09.74	&	$-$47:29:01.8	& 1.57 &	2.66	&	\\ %N/N/Y/Y	&		\\
	&		&		&	r2	&	13:27:09.93	&	$-$47:29:07.2	& 2.22 &	5.28	&	\\ %N/N/Y/Y	&		\\
 \hline
\multicolumn{9}{p{15.5cm}}{{\it Notes:} $^a$ Radio positions of the MSPs in \oc. The positions of MSPs A--E were obtained from \citet{Dai2023}, whereas others' positions were published by \citet{Chen2023}. The numbers in parentheses denote uncertainties. Positions labelled with asterisks denote the positions of the centres of the beams where the corresponding MSPs have the brightest detections \citep{Chen2023}.} \\
\multicolumn{9}{p{15.5cm}}{$^b$ Naming system for the X-ray counterparts in this work.} \\
\multicolumn{9}{p{15.5cm}}{$^c$ Positions generated by the X-ray source detection algorithm {\tt wavdetect}.} \\
\multicolumn{9}{p{15.5cm}}{$^d$ X-ray positional uncertainties at a 95\% confidence level, computed using Equation~12 in \citet{Kim2007}. } \\
\multicolumn{9}{p{15.5cm}}{$^e$ Offsets between the radio positions and X-ray positions in arcseconds.} \\
% \multicolumn{9}{p{16.5cm}}{$^f$ Indications of successful spectral extractions in each observation (Obs IDs 653/1519/13726/13727). Y for successful extractions, and N for no extraction. } \\
\multicolumn{9}{p{15.5cm}}{$^f$ Existing identifications of the X-ray sources in \oc from \citet{Haggard2009} and/or \citet{Henleywillis2018}. } \\
    \end{tabular}
    \label{tab:properties}
\end{table*}

\section{Data analysis and results}
\label{sec:results}

We performed X-ray spectral analysis using {\sc sherpa} \citep[version 4.15.0;][]{Freeman2001,Doe2007,Burke2022}, {\sc ciao}'s modelling and fitting package.

\subsection{Comparison with existing catalogues}

\oc has been studied in X-rays by several previous works \citep[e.g.,][]{Haggard2009,Henleywillis2018}. 
In particular, \citet{Henleywillis2018} performed a deep X-ray survey and catalogued 233 X-ray sources in \oc, %with available optical identifications provided. 
including information from previous optical identification campaigns \citep{Cool2013} and including some %limited 
additional counterpart identification. 
Given that %the majority of MSPs are not expected to have optical counterparts, except for those coupled with white dwarfs 
MSP optical counterparts tend to be hard to detect 
\citep[e.g., J1641+3627F in M13; see][]{Cadelano2020,Zhao2021},
we compared the X-ray %counterparts in our work 
sources that we identify as potential MSPs 
with %those 
X-ray counterpart identifications 
in \citet{Henleywillis2018}, to see if some X-ray sources have  catalogued optical counterparts, that may %rule them out as X-ray counterparts to MSPs. 
indicate the X-ray source is not the counterpart to an MSP. 
Of the 14 X-ray sources detected in this work, seven were also identified by \citet{Haggard2009} and/or \citet{Henleywillis2018} (as noted in Table~\ref{tab:properties}), %whereas 
but 
only %MSPs G and K 
X-ray sources g1 and k1 
have suggested optical counterparts. 

The optical counterpart to 
the X-ray source near 
MSP G was considered to be an anomalous subgiant or giant based on its location in the colour-magnitude diagrams \citep{Cool2013}. 
However, the timing solution of MSP G indicates it lies in a non-eclipsing black widow system with a minimum companion mass of 0.018 M$_{\sun}$ \citep{Chen2023}. %Hence, the optical counterpart to MSP G is unlikely to be genuine.
Thus, either the X-ray source is not associated with the (sub)giant; or %MSP G is not associated with the X-ray source.
the X-ray source is not associated with MSP G. 

More interestingly, two potential optical counterparts were found around the position of MSP K; one was found to be an H$\alpha$-bright source without measurable blue excess; the other one, on the contrary, was a blue source with no measurable H$\alpha$ excess \citep{Cool2013}. 
%On the other hand, 
MSP K is %found to be 
an eclipsing black widow with a minimum companion mass of 0.043 M$_{\sun}$ \citep{Chen2023}.
\citet{Cadelano2015} found the black widows M5 C \citep{Pallanca2014} and M71 A share similar light curves and positions in the colour-magnitude diagram (CMD), which are between the main sequence and the white dwarf cooling sequence \citep[see Figure 10 in][]{Cadelano2015}. 
If these are common properties for BW companions, comparing with the positions of the optical counterparts to MSP K in its CMD \citep[see Figure 3 in][]{Cool2013}, we suggest the blue source without H$\alpha$ excess is more plausibly the optical counterpart to MSP K. 
Alternatively, the other optical counterpart might be the real one to MSP K, while it is also possible that both optical counterparts are chance coincidences. 
A radio timing position for MSP K could provide sufficient localisation to precisely identify if either optical counterpart is genuine. 

We also attempted to search for possible orbital modulations for the potential X-ray counterparts to BWs in \oc\ (MSPs B, G, H, K, and L). Note that among these five BWs, we can define phases only for MSP B, as it has a published timing solution. We used the known orbital periods from radio observations \citep{Chen2023,Dai2023}, and the \chandra\ observation with the longest exposure time 
(173.7 ks; Obs. ID 13726),  which covers 12 to 22 orbits depending on the MSPs' orbital periods, and then applied {\tt pfold} script\footnote{\url{https://cxc.cfa.harvard.edu/ciao/threads/phase_bin/}} to create their orbital light curves. We found no statistically significant orbital variability for any of these counterparts, possibly due to their X-ray faintness, though the counterparts to MSPs K and L  show suggestions of X-ray modulations in their light curves. 
We suggest re-examining their X-ray light curves when timing solutions are available.

We identified more X-ray sources (seven, or a factor of 2) than \citet{Henleywillis2018} did in their work in the vicinities of the MSPs, which is likely a consequence of two factors. First, we used all four \chandra observations of \oc with a total exposure time of 290.9 ks to search for X-ray sources, whereas \citet{Henleywillis2018} only utilised 
the later 
 two \chandra observations (Obs. IDs 13726 and 13727) with a total exposure time of 222.3 ks. It is therefore expected to identify more X-ray sources, especially faint ones, with deeper observations. Also, the parameters used for {\tt wavdetect} were more conservative in \citet{Henleywillis2018}. For example, we applied a significance threshold of 10$^{-5}$, an order of magnitude higher than that used in their work, which increased the chance of detecting strong background fluctuations as sources. Since we are only investigating smaller patches of sky, the increased false-alarm rate is acceptable. Specifically, given the very precise timing positions of MSPs A--E, no search area is required to identify their counterparts. 
%Therefore, 
The total search area for the remaining MSPs is approximately 3076 arcsec$^2$. And the number of 0.492" pixels we search 
is about $1.3\times10^4$, giving a $\sim$13\% chance of finding one spurious {\tt wavdetect} detection among all 18 boxes.

We also estimated the number of chance coincidences of MSPs with unrelated X-ray sources, %which can be approximately obtained from
using 
\begin{equation}
    N_c = N_{\rm core} \times \frac{A_{\rm search}}{A_{\rm core}},
\end{equation}
where $N_c$ is the number of chance coincidences of X-ray sources, $N_{\rm core}$ is the total number of X-ray sources detected in the core region, $A_{\rm search}$ and $A_{\rm core}$ are the area of all search regions and of the core region, respectively. 
$A_{\rm search}$ is about 3076 arcsec$^2$ (see above), while 
on the other hand, a total of 81 X-ray sources were detected in the 2.37-arcmin-radius core region. 
The expected number of chance coincidences therefore yields $N_c\approx4$. Chance coincidences are more likely in the group two sources with larger error circles.

\subsection{Spectral analysis}
\label{subsec:spectra}

We performed spectral fitting for these potential X-ray counterparts. 
We first applied the spectral analysis software {\sc bxa} \citep{Buchner2014},  which connects the nested sampling algorithm UltraNest \citep{Buchner2021} with {\sc sherpa}, to generate Bayesian parameter estimation and model comparison. 
We considered four absorbed spectral models:
powerlaw (PL; {\tt xspegpwrlw} model), blackbody (BB; {\tt xsbbodyrad} model), NS hydrogen atmosphere \citep[NSA; {\tt xsnsatmos} model;][]{Heinke2006}, and %diffuse gas 
thermal plasma 
emission (APEC; {\tt xsapec} model). 
Specifically for the {\tt xsnsatmos} model, the parameters of NS mass and radius were fixed to 1.4 M$_{\sun}$ and 10 km, respectively, and the distance to \oc was assumed to be 5.43 kpc \citep{Baumgardt2021}. 
PL, BB, and/or NSA spectra are commonly observed from MSPs in GCs \citep[see e.g.,][]{Zhao2022}, while PL models generally describe AGN spectra well, and APEC models generally describe the spectra from coronally active binaries. 
%and APEC spectra are usually seen from general X-ray sources. 
We also considered two combined models, PL+BB and PL+NSA, as some MSPs
(typically spider pulsars)
show both non-thermal and thermal components in their spectra, e.g. PSR J0437$-$4715 \citep{Zavlin02}, and 47 Tuc J and W \citep{Bogdanov2006}.
The interstellar X-ray absorption was modelled using {\tt xstbabs}, with the {\it wilm} elemental abundance \citep{Wilms2000} and {\it vern} photoelectric absorption cross section \citep{Verner1996}. 
The hydrogen column number density ($N_{\rm H}$) towards \oc was assumed to be fixed at $N_{\rm H}=1.05\times10^{21}$~cm$^{-2}$, 
%This $N_{\rm H}$ was
estimated from the interstellar reddening $E(B-V)$ towards \oc \citep[2010 edition]{Harris1996}, and the correlation between the optical extinction $A_V$ and $N_{\rm H}$ \citep{Bahramian2015}, using the ratio $A_V/E(B-V)=3.1$ \citep{Cardelli1989}. 
We adopted uniform priors for the PL photon index ($\Gamma$)  between $-$3 and 10 and $\log{T_{\rm eff}}$ of NSA (where $T_{\rm eff}$ is the unredshifted effective temperature of the NS surface in units of Kelvin) between 5 and 6.5, and log-uniform priors for $kT_{\rm BB}$ of BB between 0.0001 and 10 and $kT_{\rm APEC}$ of APEC between 0.008 and 30, where $kT_{\rm BB}$ and $kT_{\rm APEC}$ are unredshifted blackbody temperature and plasma temperature, respectively, both in units of keV.
All priors for normalisations were set to log-uniform from 10$^{-8}$ to 10$^{2}$. 
X-ray data were  grouped to at least one photon per bin for {\sc bxa} fitting.
We used the {\tt model\_compare} script implemented in {\sc bxa} to compare those models for each source by comparing the Bayesian evidences. Specifically, we applied Jeffreys' scale \citep{Jeffreys1939} and ruled out models if $\Delta\log_{10}Z = \log_{10}Z_i - \log_{10}Z_0 < -1$, where $Z_0$ represents the highest evidence among all models for a source, and $Z_i$ is the evidence of each model. 

We summarise the spectral fitting and model comparison results obtained using {\sc bxa} in Table~\ref{tab:model_compare}. Except for h2, for which the X-ray spectrum only favours the APEC model, the X-ray spectra of all sources can be well fitted by more than one spectral model. 
Thus, to further examine the predominant component (thermal or non-thermal) in the X-ray emission of each source, we fitted their spectra with PL+BB and PL+NSA models by fixing %the photon indices of 
$\Gamma=2$. 
We computed the fractions of flux contributed by the BB and NSA components relative to the total fluxes in three bands (0.5--2 keV, 2--8 keV, and 0.5--10 keV, respectively), and listed the results in Table~\ref{tab:thermal_frac}.
In most PL+NSA fits, either the PL or NSA can dominate, except h1 and k1 where the PL dominates, and r1 where the NSA dominates. PL+BB fits, on the other hand, require PL domination for all but c1, e1, and r1.     
Based on the results in Table~\ref{tab:model_compare} and Table~\ref{tab:thermal_frac}, as well as previous studies about typical X-ray emission from different types of MSPs \citep[see e.g.,][]{Zhao2022,Lee2023}, 
we selected our preferred spectral model for each source
(see Table~\ref{tab:X-ray_properties}; other useful X-ray properties also provided).

We identify three sources (h2, r1, r2) as not matching our spectral assumptions for MSPs--which are that an MSP should have a spectrum dominated by a power-law of photon index 1--2, and/or an NSA component consistent with emission from 
a fraction of a percent of a NS surface. 
%a few percent of a NS surface.
For instance, the inferred NSATMOS hot spot radii for MSPs in 47 Tuc range from 0.28 to 1.7 km \citep{Bogdanov2006}, thus giving emitting fractions of a NS surface of 0.014\% to 0.5\% of the surface area of a 12 km NS.

%Specifically, 
The X-ray spectrum of h2 can only be well fitted by an APEC model, whereas all other five models are ruled out, based on the comparison of Bayesian evidences (the APEC model is more than three orders of magnitude favoured than other models).
Hence, the X-ray emission from h2 likely originates from hot plasma (i.e., either coronal activity in an active binary, or a cataclysmic variable) rather than an MSP.
A (foreground) active binary would likely have a high optical/X-ray flux ratio \citep{Verbunt08}, suggesting that the optical counterpart may be bright. Indeed, we found that there is a {\it Gaia} DR3 source \citep[Gmag=15.1;][]{GaiaCollaboration2023} close to the X-ray position of h2, with an angular separation of $\sim$0.3 arcsec. 
Note that the X-ray positional uncertainty of h2 is 0.22 arcsec at 95\% confidence level (Table~\ref{tab:properties}).
However, due to the lack of proper motions and parallax of that {\it Gaia} source, it is hard to determine if it is a member of \oc and/or an optical counterpart to h2.
Follow-up optical observations are required to clarify the nature of that potential optical counterpart. Additionally, obtaining a precise timing position for MSP H is crucial to definitively determine its %corresponding 
X-ray counterpart.

\begin{figure*}
    \centering
    \includegraphics[width=0.425\textwidth]{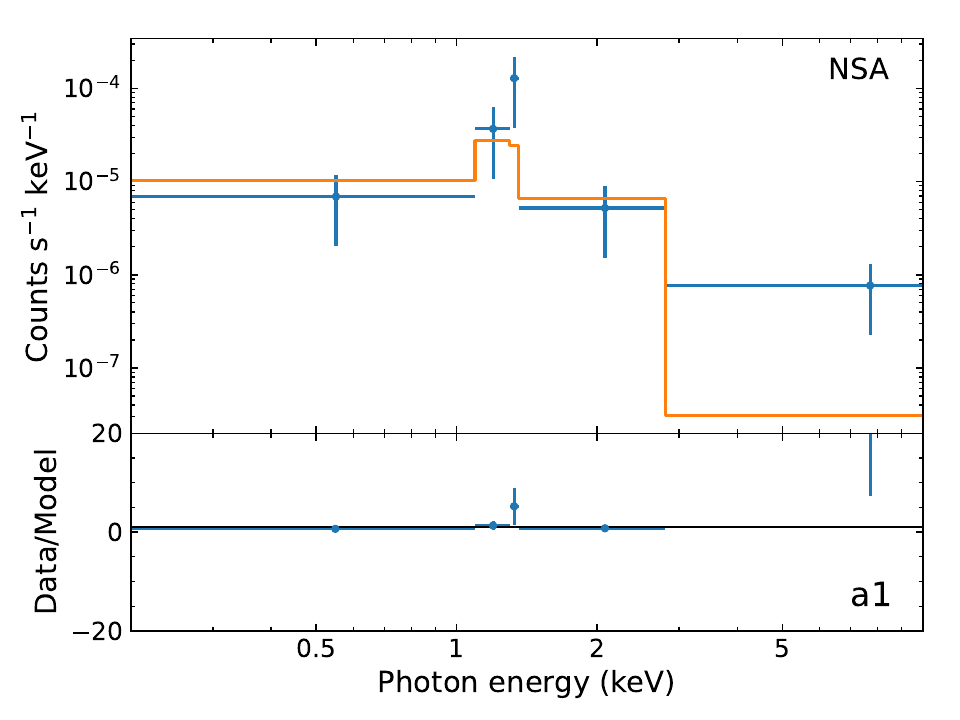}
    \includegraphics[width=0.425\textwidth]{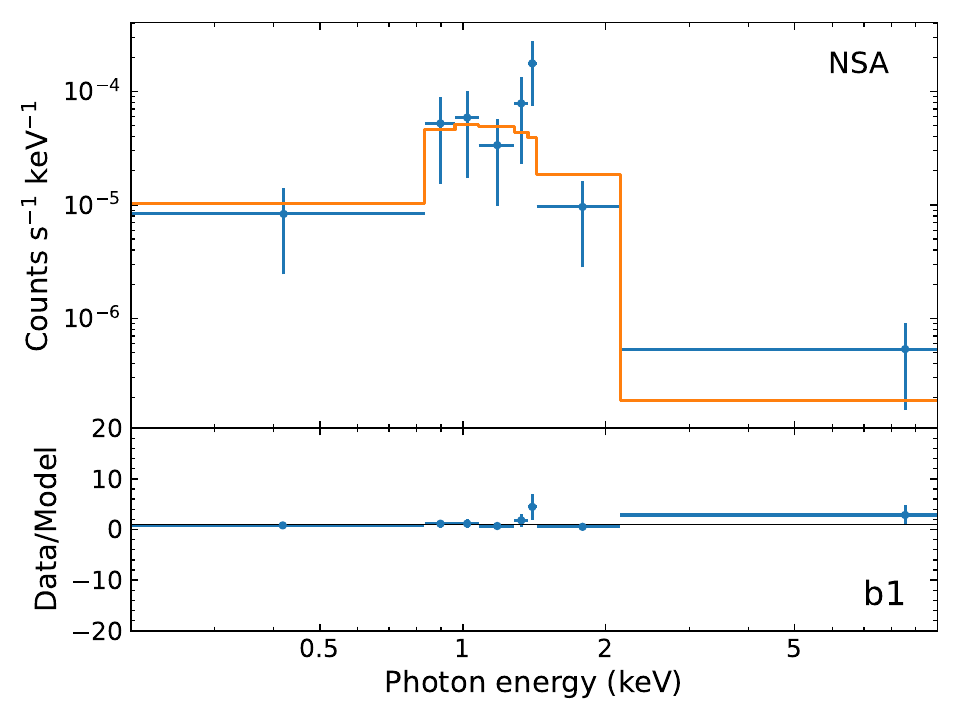}
    \includegraphics[width=0.425\textwidth]{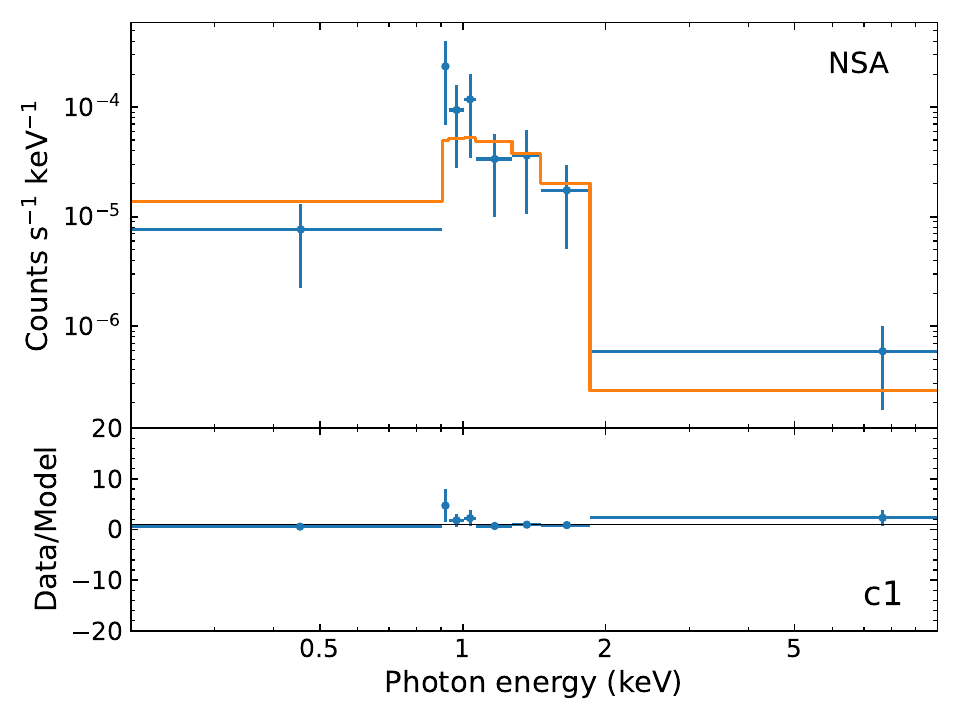}
    \includegraphics[width=0.425\textwidth]{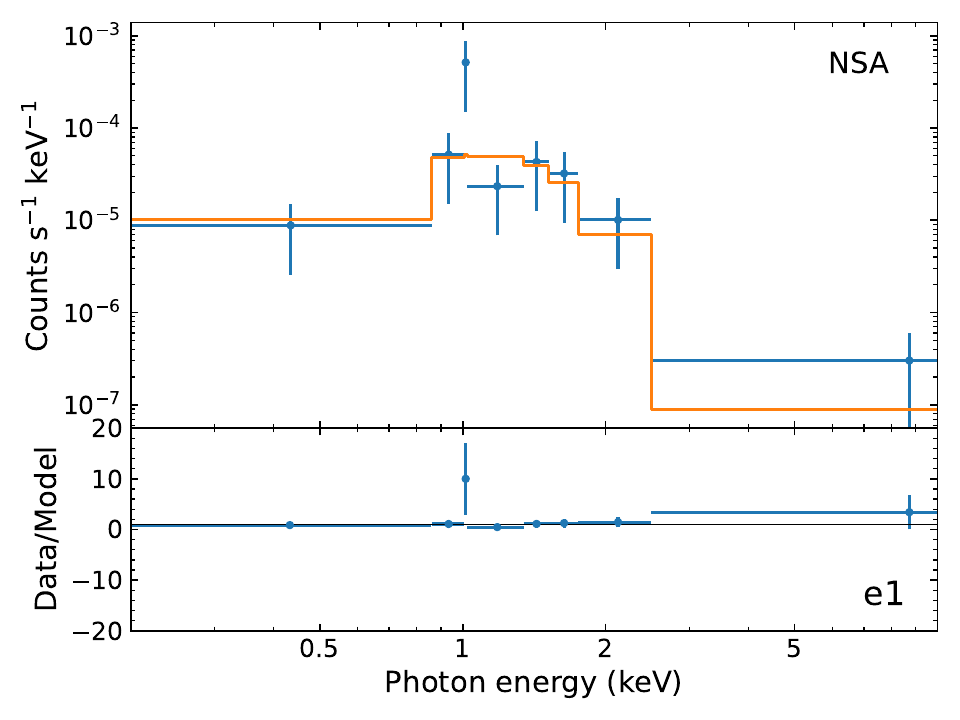}
    \includegraphics[width=0.425\textwidth]{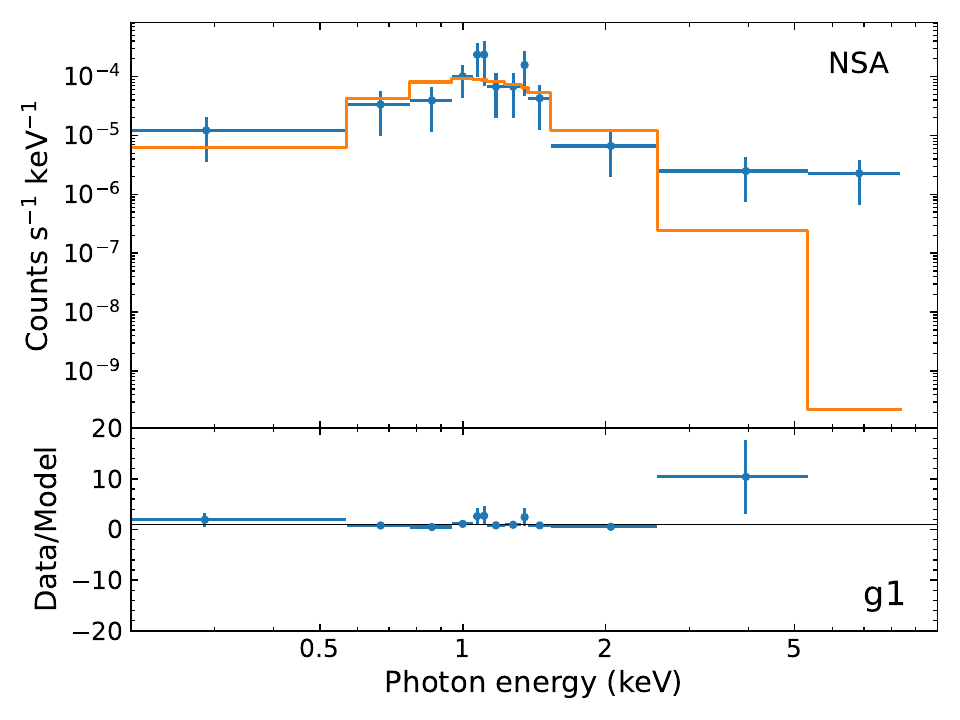}
    \includegraphics[width=0.425\textwidth]{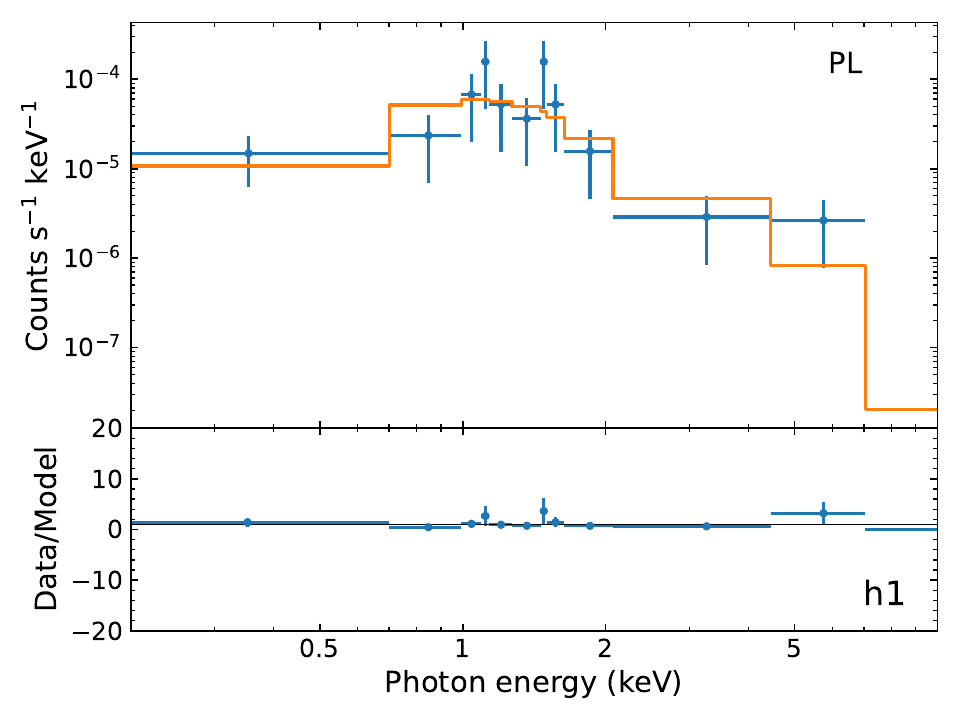}
    \includegraphics[width=0.425\textwidth]{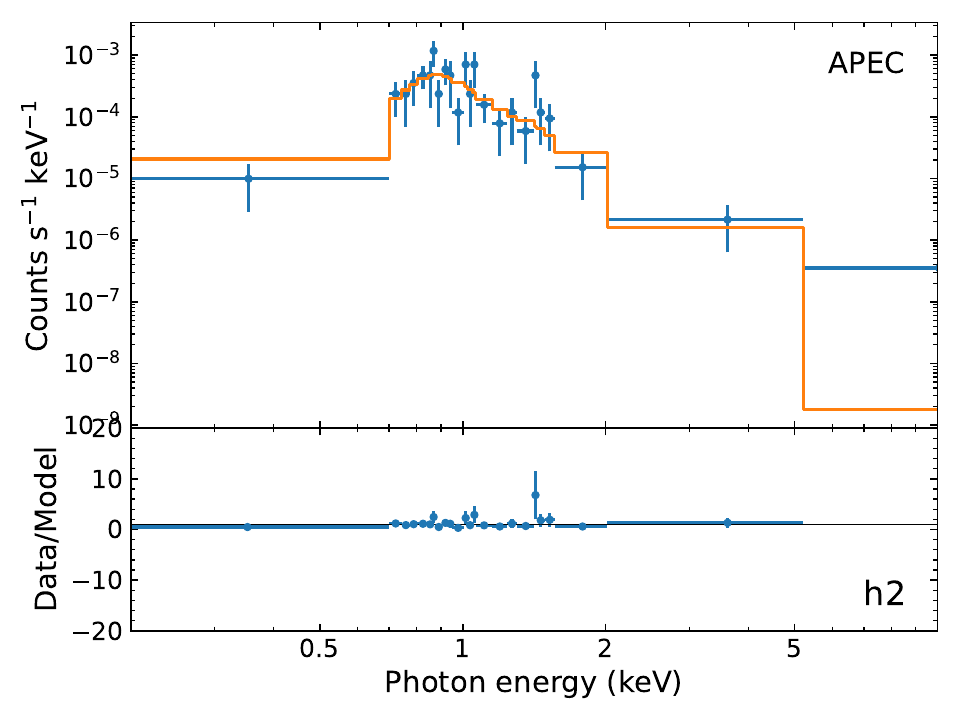}
    \includegraphics[width=0.425\textwidth]{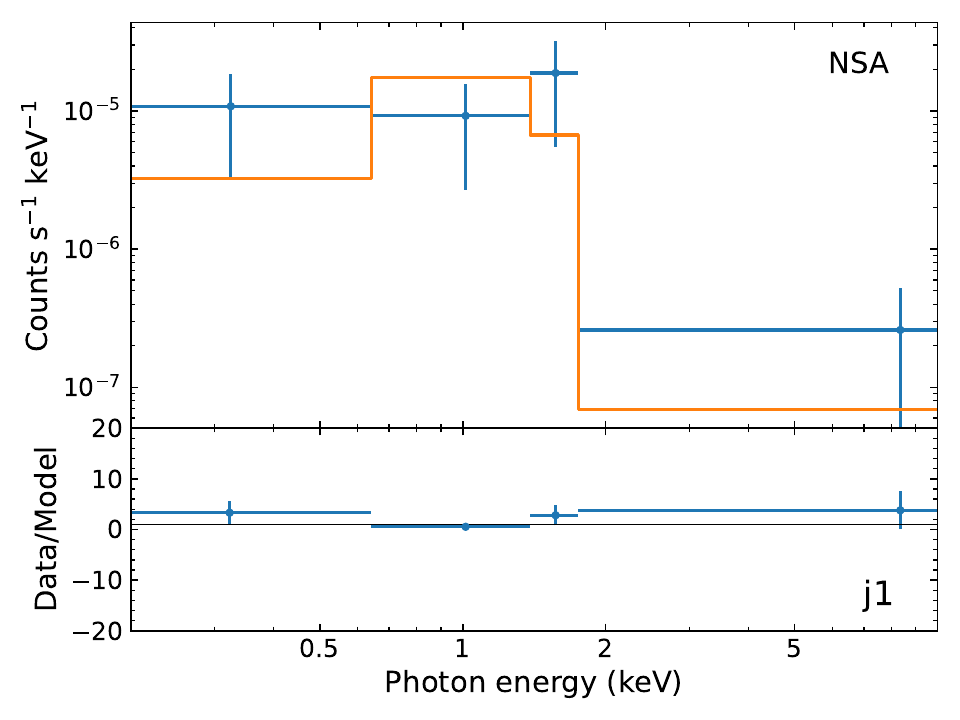}
    \caption{X-ray spectra and the best fits of eight counterparts in the energy range 0.3--8 keV (see Table~\ref{tab:X-ray_properties} for the best-fit model of each source). Data are grouped to a signal-to-noise ratio of at least 1 per bin for display purpose, while data were grouped to at least 1 photon per bin for fitting processes. The spectra were fitted with WSTAT statistics via {\sc bxa}. The names of each source and the corresponding best-fit model are labeled in the lower-right corner and the upper-right corner, respectively, of each figure. }
    \label{fig:spectra1}
\end{figure*}

\begin{figure*}
    \centering
    \includegraphics[width=0.425\textwidth]{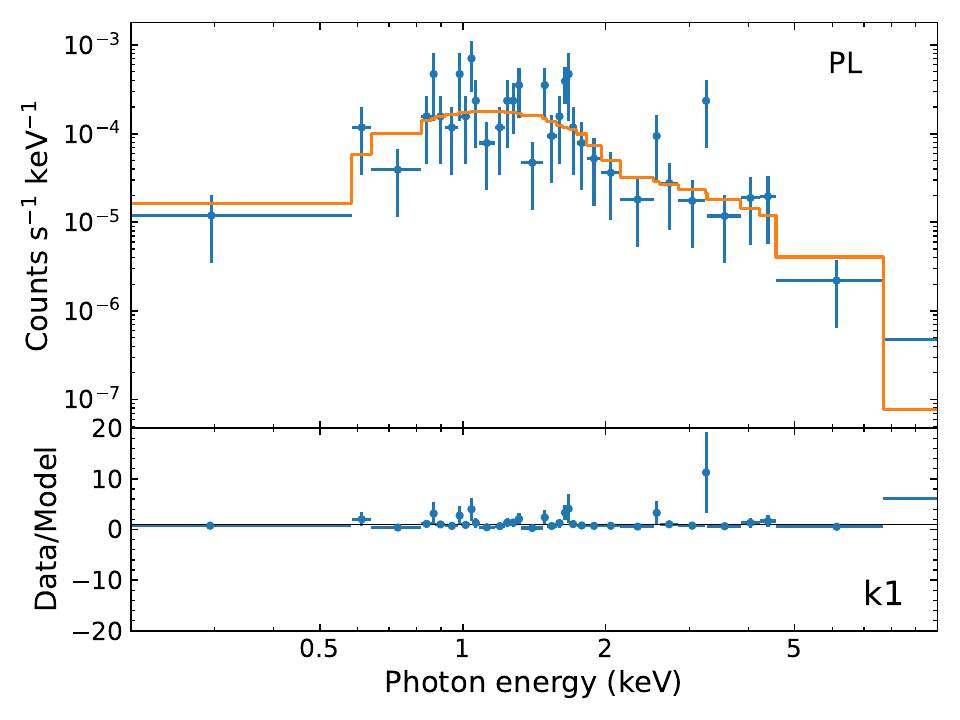}
    \includegraphics[width=0.425\textwidth]{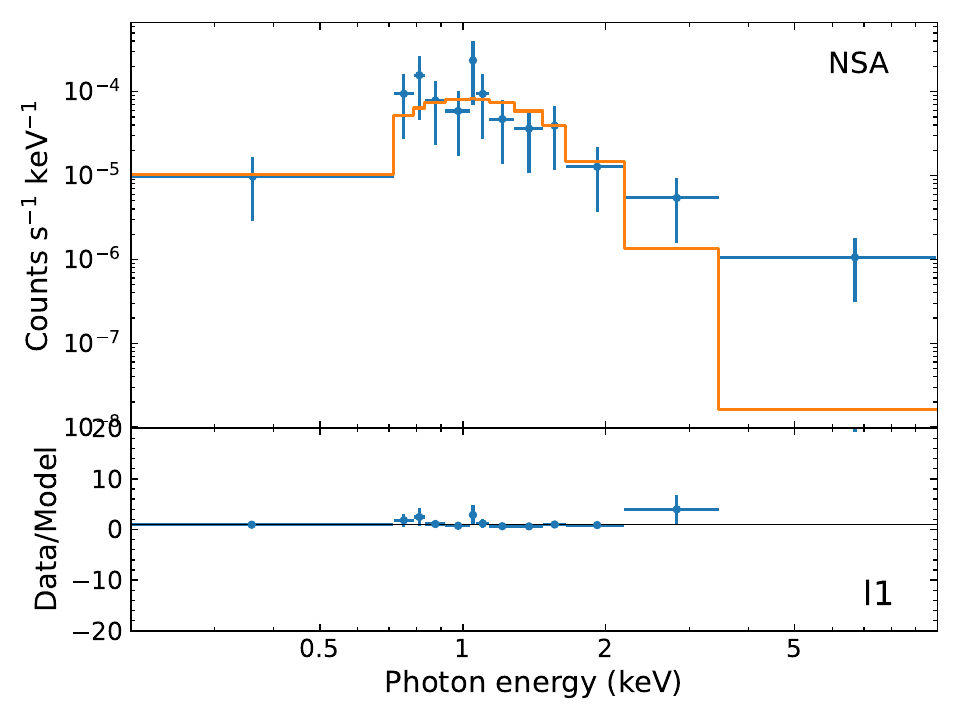}
    \includegraphics[width=0.425\textwidth]{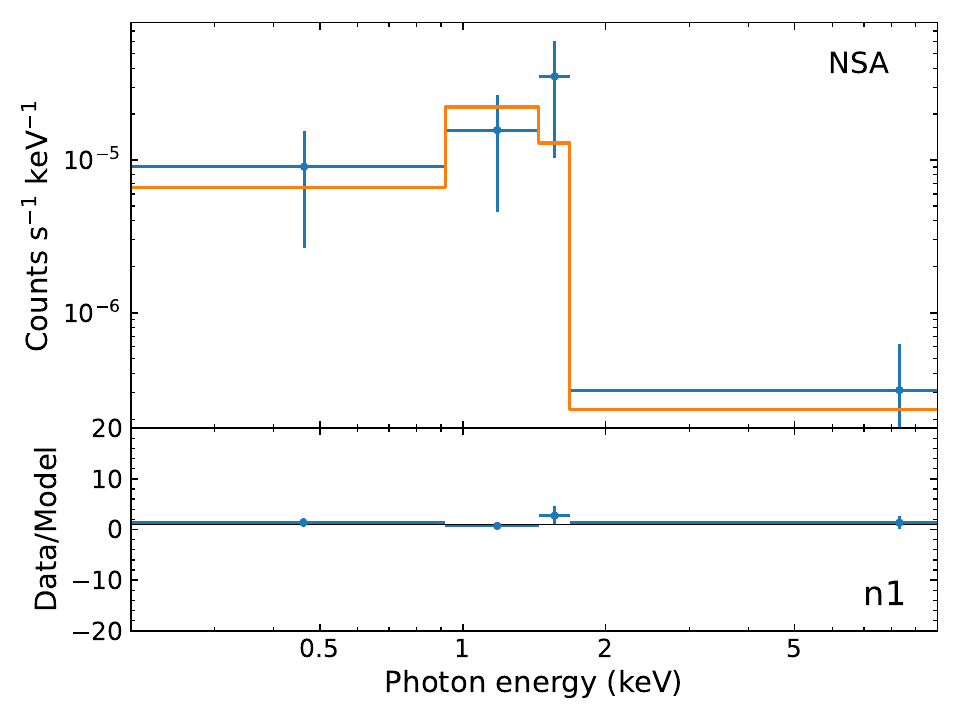}
    \includegraphics[width=0.425\textwidth]{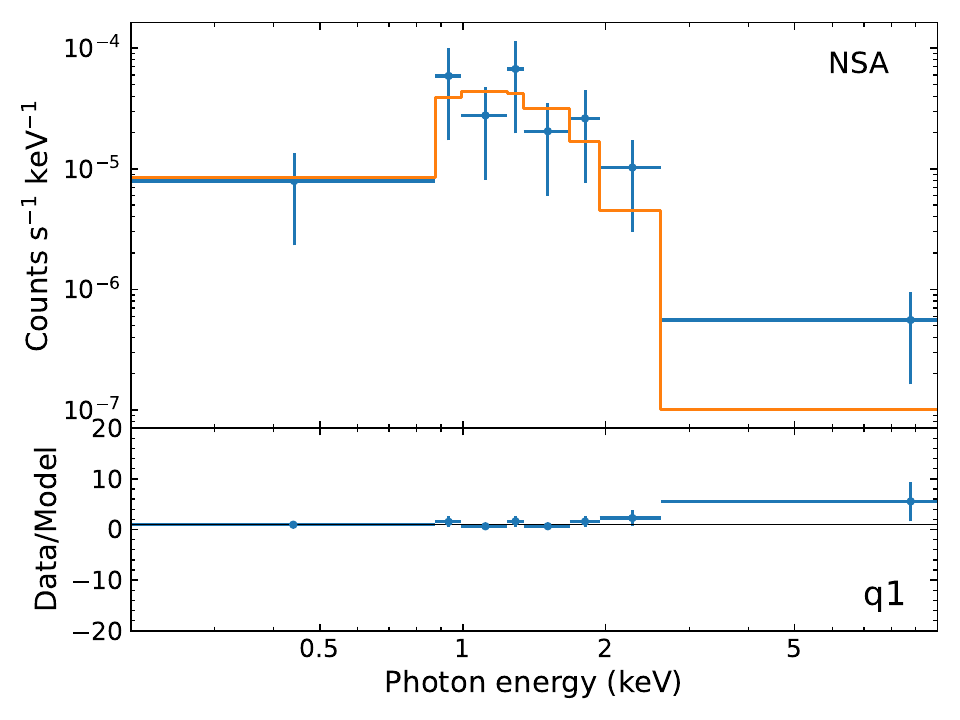}
    \includegraphics[width=0.425\textwidth]{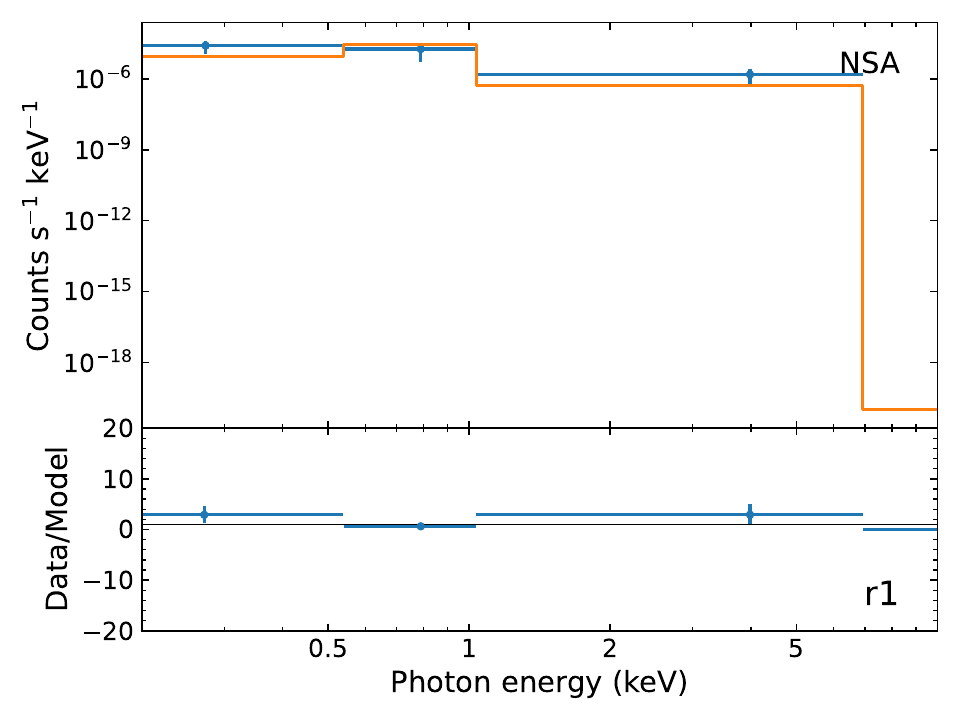}
    \includegraphics[width=0.425\textwidth]{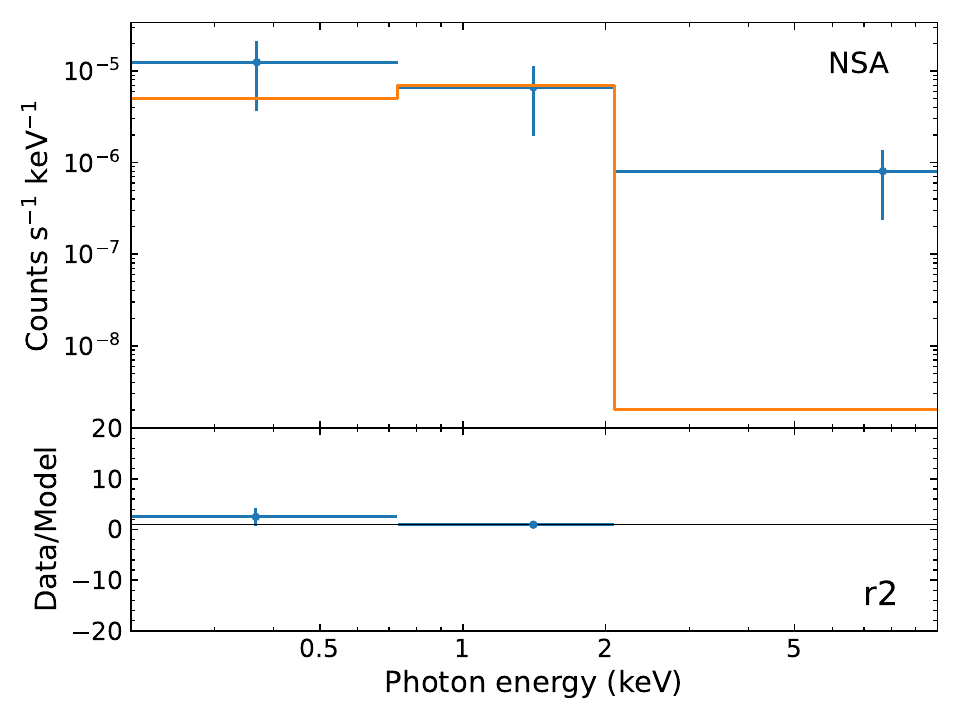}
    \caption{X-ray spectra and the best fits of six counterparts in the energy range 0.3--8 keV. See the caption of Figure~\ref{fig:spectra1} for details.}
    \label{fig:spectra2}
\end{figure*}

For r1 and r2, the fitted fractions of the NS surface emitting from the NSA model are $12.29^{+44.27}_{-11.34}$ and $0.15^{+9.73}_{-0.15}$, respectively.
Thus, the NSA component for r1 needs a much larger emitting area than the full neutron star surface to reproduce its spectrum, ruling out r1.
On the other hand, given the large uncertainty of the emission fraction of r2, we cannot completely rule out r2 as a MSP counterpart.
However, due to its poorly extracted spectrum and large fitting uncertainties, we do not consider r2 a confident X-ray counterpart. 
A future precise timing solution of MSP R will eventually determine its true X-ray counterpart. 
We  show the X-ray spectra and the best fits for those counterparts in Figure~\ref{fig:spectra1} and Figure~\ref{fig:spectra2}.

Additionally, despite the non-detection of X-ray counterpart to MSP D, we are able to constrain its X-ray luminosity given its precise timing position. We first calculate the count rate and flux in 0.3--8 keV of a 1-arcsec-radius region centred at MSP D for each observation using {\tt srcflux}.
By assuming a PL model with a fixed photon index $\Gamma=2$, 
the 1-$\sigma$ upper limit of the unabsorbed X-ray luminosity of MSP D was estimated to be 4.0$\times$10$^{30}$ \ergs, at a distance of 5.43 kpc. 

\begin{landscape}

\begin{table}
    \centering
    \caption{Spectral fitting and model comparison results from {\sc bxa}}
    \begin{tabular}{lcccccccccccccccccccc}
\hline
Name & & \multicolumn{2}{c}{APEC} & & \multicolumn{2}{c}{BB} & & \multicolumn{2}{c}{NSA} & & \multicolumn{2}{c}{PL} & & \multicolumn{3}{c}{PL+BB} & & \multicolumn{3}{c}{PL+NSA}\\
\cline{3-4} \cline{6-7} \cline{9-10} \cline{12-13} \cline{15-17} \cline{19-21}
 & & $kT_{\rm APEC}$ & $\Delta\log_{10}{Z}$ & & $kT_{\rm BB}$ & $\Delta\log_{10}{Z}$ & & $\log{T_{\rm eff}}$ & $\Delta\log_{10}{Z}$ & & $\Gamma$ & $\Delta\log_{10}{Z}$ & & $\Gamma$ & $kT_{\rm BB}$ & $\Delta\log_{10}{Z}$ & & $\Gamma$ & $\log{T_{\rm eff}}$ & $\Delta\log_{10}{Z}$\\
\hline
a1	& &	2.23$^{+4.30}_{-0.88}$	&	{\bf $-$0.26}	& &	0.30$^{+0.13}_{-0.08}$	&	{\bf $-$0.41}	& &	6.29$^{+0.14}_{-0.18}$	&	{\bf 0.00}	& &	2.65$^{+0.85}_{-0.76}$	&	{\bf $-$0.69}	& &	2.72$^{+3.09}_{-1.59}$ & 0.20$^{+0.18}_{-0.20}$	&	{\bf $-$0.45}	& &	2.75$^{+4.08}_{-3.48}$ & 6.22$^{+0.19}_{-0.53}$	&	{\bf $-$0.16}	\\
b1	& &	2.13$^{+2.62}_{-0.70}$	&	{\bf $-$0.36}	& &	0.32$^{+0.08}_{-0.06}$	&	{\bf $-$0.59}	& &	6.35$^{+0.10}_{-0.12}$	&	{\bf 0.00}	& &	2.51$^{+0.54}_{-0.50}$	&	{\bf $-$0.61}	& &	2.59$^{+2.86}_{-0.90}$ & 0.21$^{+0.18}_{-0.21}$	&	{\bf $-$0.39}	& &	2.67$^{+4.17}_{-2.86}$ & 6.29$^{+0.14}_{-0.63}$	&	{\bf $-$0.19}	\\
c1	& &	1.12$^{+0.26}_{-0.20}$	&	{\bf 0.00}	& &	0.25$^{+0.06}_{-0.05}$	&	{\bf $-$0.89}	& &	6.22$^{+0.14}_{-0.16}$	&	{\bf $-$0.16}	& &	3.09$^{+0.60}_{-0.58}$	&	{\bf $-$0.96}	& &	3.16$^{+2.72}_{-1.38}$ & 0.18$^{+0.12}_{-0.18}$	&	{\bf $-$0.78}	& &	3.07$^{+3.89}_{-3.53}$ & 6.18$^{+0.16}_{-0.23}$	&	{\bf $-$0.32}	\\
e1	& &	1.47$^{+0.94}_{-0.35}$	&	{\bf $-$0.24}	& &	0.30$^{+0.09}_{-0.06}$	&	{\bf $-$0.63}	& &	6.30$^{+0.13}_{-0.14}$	&	{\bf 0.00}	& &	2.87$^{+0.56}_{-0.55}$	&	{\bf $-$0.55}	& &	2.99$^{+1.99}_{-0.98}$ & 0.11$^{+0.24}_{-0.11}$	&	{\bf $-$0.45}	& &	2.87$^{+4.25}_{-3.18}$ & 6.25$^{+0.16}_{-0.46}$	&	{\bf $-$0.53}	\\
g1	& &	1.71$^{+0.64}_{-0.34}$	&	{\bf $-$0.06}	& &	0.24$^{+0.05}_{-0.04}$	&	$-$0.99	& &	6.18$^{+0.13}_{-0.12}$	&	{\bf $-$0.11}	& &	2.86$^{+0.49}_{-0.45}$	&	{\bf $-$0.23}	& &	2.81$^{+0.62}_{-1.70}$ & 0.04$^{+0.25}_{-0.04}$	&	{\bf $-$0.13}	& &	1.72$^{+1.82}_{-3.71}$ & 6.10$^{+0.17}_{-0.48}$	&	{\bf 0.00}	\\
h1	& &	2.78$^{+2.54}_{-0.86}$	&	{\bf 0.00}	& &	0.34$^{+0.10}_{-0.07}$	&	$-$1.38	& &	6.37$^{+0.09}_{-0.13}$	&	{\bf $-$0.67}	& &	2.47$^{+0.48}_{-0.43}$	&	{\bf $-$0.08}	& &	2.46$^{+0.55}_{-0.45}$ & 0.01$^{+0.32}_{-0.01}$	&	{\bf $-$0.11}	& &	2.41$^{+0.71}_{-0.77}$ & 5.83$^{+0.52}_{-0.59}$	&	{\bf $-$0.11}	\\
h2	& &	0.76$^{+0.08}_{-0.10}$	&	{\bf 0.00}	& &	0.18$^{+0.02}_{-0.01}$	&	$-$3.74	& &	6.02$^{+0.08}_{-0.07}$	&	$-$3.43	& &	3.63$^{+0.29}_{-0.27}$	&	$-$6.39	& &	3.01$^{+4.77}_{-4.03}$ & 0.18$^{+0.02}_{-0.01}$	&	$-$3.93	& &	2.49$^{+4.74}_{-3.76}$ & 6.02$^{+0.07}_{-0.07}$	&	$-$3.61	\\
j1	& &	3.22$^{+7.50}_{-1.85}$	&	{\bf $-$0.73}	& &	0.29$^{+0.17}_{-0.11}$	&	$-$1.21	& &	6.09$^{+0.26}_{-0.30}$	&	{\bf $-$0.34}	& &	3.72$^{+1.36}_{-1.13}$	&	{\bf $-$0.23}	& &	3.87$^{+2.05}_{-1.29}$ & 0.02$^{+0.39}_{-0.02}$	&	{\bf $-$0.36}	& &	3.96$^{+3.44}_{-2.02}$ & 5.87$^{+0.45}_{-0.58}$	&	{\bf 0.00}	\\
k1	& &	3.92$^{+2.13}_{-1.06}$	&	{\bf 0.00}	& &	0.55$^{+0.06}_{-0.05}$	&	$-$3.23	& &	6.49$^{+0.01}_{-0.02}$	&	$-$4.67	& &	1.99$^{+0.22}_{-0.22}$	&	{\bf $-$0.36}	& &	1.99$^{+0.23}_{-0.23}$ & 0.01$^{+0.28}_{-0.01}$	&	{\bf $-$0.37}	& &	1.97$^{+0.23}_{-0.23}$ & 5.63$^{+0.56}_{-0.44}$	&	{\bf $-$0.55}	\\
l1	& &	1.20$^{+0.86}_{-0.41}$	&	{\bf $-$0.42}	& &	0.25$^{+0.06}_{-0.05}$	&	$-$1.13	& &	6.22$^{+0.13}_{-0.13}$	&	{\bf $-$0.07}	& &	2.97$^{+0.47}_{-0.42}$	&	{\bf 0.00}	& &	2.98$^{+0.54}_{-0.50}$ & 0.01$^{+0.26}_{-0.01}$	&	{\bf $-$0.10}	& &	2.91$^{+1.01}_{-1.97}$ & 6.02$^{+0.27}_{-0.69}$	&	{\bf $-$0.38}	\\
n1	& &	3.08$^{+6.49}_{-1.60}$	&	{\bf $-$0.39}	& &	0.32$^{+0.12}_{-0.09}$	&	{\bf $-$0.48}	& &	6.27$^{+0.15}_{-0.20}$	&	{\bf 0.00}	& &	2.91$^{+0.87}_{-0.86}$	&	{\bf $-$0.41}	& &	3.13$^{+1.96}_{-1.18}$ & 0.07$^{+0.33}_{-0.07}$	&	{\bf $-$0.21}	& &	3.10$^{+3.61}_{-2.70}$ & 6.18$^{+0.22}_{-0.71}$	&	{\bf $-$0.07}	\\
q1	& &	2.80$^{+4.39}_{-1.31}$	&	{\bf $-$0.19}	& &	0.37$^{+0.12}_{-0.08}$	&	{\bf $-$0.56}	& &	6.38$^{+0.09}_{-0.14}$	&	{\bf 0.00}	& &	2.50$^{+0.54}_{-0.54}$	&	{\bf $-$0.45}	& &	2.62$^{+2.94}_{-1.02}$ & 0.15$^{+0.28}_{-0.15}$	&	{\bf $-$0.43}	& &	2.55$^{+3.85}_{-2.44}$ & 6.31$^{+0.13}_{-0.81}$	&	{\bf $-$0.19}	\\
r1	& &	0.06$^{+0.05}_{-0.02}$	&	{\bf $-$0.72}	& &	0.08$^{+0.02}_{-0.01}$	&	$-$1.69	& &	5.52$^{+0.15}_{-0.08}$	&	{\bf $-$0.89}	& &	7.81$^{+1.33}_{-1.46}$	&	{\bf 0.00}	& &	7.76$^{+1.43}_{-1.59}$ & 0.01$^{+0.34}_{-0.01}$	&	{\bf $-$0.11}	& &	7.76$^{+1.48}_{-1.64}$ & 5.58$^{+0.50}_{-0.36}$	&	{\bf $-$0.02}	\\
r2	& &	0.49$^{+6.11}_{-0.38}$	&	{\bf $-$0.62}	& &	0.14$^{+0.20}_{-0.05}$	&	{\bf $-$0.99}	& &	5.76$^{+0.40}_{-0.27}$	&	{\bf $-$0.37}	& &	5.95$^{+2.26}_{-2.30}$	&	{\bf $-$0.17}	& &	6.01$^{+2.18}_{-2.74}$ & 0.02$^{+0.34}_{-0.02}$	&	{\bf $-$0.23}	& &	5.44$^{+2.81}_{-3.72}$ & 5.65$^{+0.48}_{-0.38}$	&	{\bf 0.00}	\\
\hline
\multicolumn{21}{p{24cm}}{{\it Notes}: Spectral fits of six models (APEC, BB, NSA, PL, PL+BB, and PL+NSA) generated by {\sc bxa} (see Section~\ref{subsec:spectra}). For each model, the primary parameter and the relative evidence ($\Delta\log_{10}{Z}$; normalised to the largest evidence in each case) are shown. Values of $\Delta\log_{10}{Z}$ in bold indicate accepted fits.
Uniform model priors are assumed, with a cut of $-1$ to rule out models.
The errors are reported at 1-$\sigma$ confidence level.}
    \end{tabular}
    \label{tab:model_compare}
\end{table}

\begin{table}
    \centering
    \caption{Fractions of thermal components}
    \begin{tabular}{lccccccccc}
\hline
Name & & \multicolumn{3}{c}{PL+BB} & & \multicolumn{3}{c}{PL+NSA} \\
\cline{3-5} \cline{7-9}
 & & $f_{\rm BB, 0.5-2}$ & $f_{\rm BB, 2-8}$ & $f_{\rm BB, 0.5-10}$ & & $f_{\rm NSA, 0.5-2}$ & $f_{\rm NSA, 2-8}$ & $f_{\rm NSA, 0.5-10}$ \\
 \hline
a1	&	& 0.00$_{-0.00}^{+0.37}$ & 0.00$_{-0.00}^{+0.14}$ & 0.00$_{-0.00}^{+0.27}$	&	& 0.03$_{-0.03}^{+0.97}$ & 0.00$_{-0.00}^{+0.98}$ & 0.02$_{-0.02}^{+0.98}$	\\
b1	&	& 0.00$_{-0.00}^{+0.02}$ & 0.00$_{-0.00}^{+0.02}$ & 0.00$_{-0.00}^{+0.06}$	&	& 0.01$_{-0.01}^{+0.99}$ & 0.00$_{-0.00}^{+0.97}$ & 0.00$_{-0.00}^{+0.99}$	\\
c1	&	& 0.00$_{-0.00}^{+1.00}$ & 0.00$_{-0.00}^{+0.98}$ & 0.00$_{-0.00}^{+1.00}$	&	& 0.98$_{-0.98}^{+0.02}$ & 0.73$_{-0.73}^{+0.27}$ & 0.96$_{-0.96}^{+0.04}$	\\
e1	&	& 0.00$_{-0.00}^{+0.75}$ & 0.00$_{-0.00}^{+0.26}$ & 0.00$_{-0.00}^{+0.63}$	&	& 0.43$_{-0.43}^{+0.57}$ & 0.02$_{-0.02}^{+0.98}$ & 0.26$_{-0.26}^{+0.74}$	\\
g1	&	& 0.00$_{-0.00}^{+0.48}$ & 0.00$_{-0.00}^{+0.03}$ & 0.00$_{-0.00}^{+0.32}$	&	& 0.37$_{-0.37}^{+0.61}$ & 0.01$_{-0.01}^{+0.68}$ & 0.22$_{-0.22}^{+0.74}$	\\
h1	&	& 0.00$_{-0.00}^{+0.00}$ & 0.00$_{-0.00}^{+0.00}$ & 0.00$_{-0.00}^{+0.00}$	&	& 0.00$_{-0.00}^{+0.17}$ & 0.00$_{-0.00}^{+0.00}$ & 0.00$_{-0.00}^{+0.09}$	\\
% h2	&	& 1.00$_{-0.04}^{+0.00}$ & 0.79$_{-0.67}^{+0.21}$ & 1.00$_{-0.08}^{+0.00}$	&	& 1.00$_{-0.02}^{+0.00}$ & 0.96$_{-0.57}^{+0.04}$ & 1.00$_{-0.05}^{+0.00}$	\\
j1	&	& 0.00$_{-0.00}^{+0.02}$ & 0.00$_{-0.00}^{+0.00}$ & 0.00$_{-0.00}^{+0.05}$	&	& 0.02$_{-0.02}^{+0.94}$ & 0.00$_{-0.00}^{+0.25}$ & 0.01$_{-0.01}^{+0.90}$	\\
k1	&	& 0.00$_{-0.00}^{+0.00}$ & 0.00$_{-0.00}^{+0.00}$ & 0.00$_{-0.00}^{+0.00}$	&	& 0.00$_{-0.00}^{+0.01}$ & 0.00$_{-0.00}^{+0.00}$ & 0.00$_{-0.00}^{+0.00}$	\\
l1	&	& 0.00$_{-0.00}^{+0.45}$ & 0.00$_{-0.00}^{+0.01}$ & 0.00$_{-0.00}^{+0.29}$	&	& 0.57$_{-0.57}^{+0.43}$ & 0.02$_{-0.02}^{+0.90}$ & 0.39$_{-0.39}^{+0.60}$	\\
n1	&	& 0.00$_{-0.00}^{+0.23}$ & 0.00$_{-0.00}^{+0.08}$ & 0.00$_{-0.00}^{+0.23}$	&	& 0.03$_{-0.03}^{+0.97}$ & 0.00$_{-0.00}^{+0.97}$ & 0.01$_{-0.01}^{+0.98}$	\\
q1	&	& 0.00$_{-0.00}^{+0.03}$ & 0.00$_{-0.00}^{+0.02}$ & 0.00$_{-0.00}^{+0.05}$	&	& 0.00$_{-0.00}^{+0.96}$ & 0.00$_{-0.00}^{+0.83}$ & 0.00$_{-0.00}^{+0.94}$	\\
r1	&	& 0.98$_{-0.43}^{+0.02}$ & 0.00$_{-0.00}^{+0.02}$ & 0.97$_{-0.59}^{+0.03}$	&	& 1.00$_{-0.11}^{+0.00}$ & 0.07$_{-0.07}^{+0.80}$ & 0.99$_{-0.20}^{+0.01}$	\\
r2	&	& 0.00$_{-0.00}^{+0.32}$ & 0.00$_{-0.00}^{+0.01}$ & 0.00$_{-0.00}^{+0.34}$	&	& 0.20$_{-0.20}^{+0.80}$ & 0.00$_{-0.00}^{+0.61}$ & 0.10$_{-0.10}^{+0.90}$	\\
\hline
\multicolumn{9}{p{11cm}}{{\it Notes}: Flux fractions of BB components (columns 2--4) and NSA components (column 5--7) over total unabsorbed fluxes in the bands 0.5--2 keV, 2--8 keV, and 0.5--10 keV, respectively. The photon indices of the PL model were fixed at 2 in all fits. The errors are reported at 1-$\sigma$ confidence level.}
    \end{tabular}

    \label{tab:thermal_frac}
\end{table}

\end{landscape}

\begin{table*}
    \centering
    \caption{X-ray properties of the counterparts}
    \begin{tabular}{lcrrrrrcc}
\hline
Name & Model$^a$ & ${F_{0.5-2}}^b$ & ${F_{2-8}}^b$ & ${F_{0.5-10}}^b$ & Xcolour$^c$ & ${L_{0.5-10}}^d$ & MSP$^e$ & Type$^f$ \\
\hline
a1 & NSA & 2.62$_{-0.89}^{+1.14}$ & 0.25$_{-0.15}^{+0.29}$ & 2.96$_{-0.98}^{+1.20}$ & 2.45$_{-0.85}^{+1.23}$ & 1.04$_{-0.34}^{+0.42}$ & A & I \\
b1 & NSA & 9.82$_{-2.37}^{+3.15}$ & 1.42$_{-0.70}^{+0.94}$ & 11.40$_{-2.93}^{+3.21}$ & 2.08$_{-0.58}^{+0.77}$ & 4.03$_{-1.03}^{+1.13}$ & B & eBW \\
c1 & NSA & 5.12$_{-1.30}^{+1.82}$ & 0.34$_{-0.21}^{+0.39}$ & 5.57$_{-1.34}^{+1.77}$ & 2.91$_{-0.88}^{+1.12}$ & 1.97$_{-0.47}^{+0.62}$ & C & I \\
e1 & NSA & 4.46$_{-1.18}^{+1.41}$ & 0.47$_{-0.25}^{+0.42}$ & 5.00$_{-1.25}^{+1.54}$ & 2.38$_{-0.76}^{+0.94}$ & 1.76$_{-0.44}^{+0.54}$ & E & I \\
g1 & NSA & 9.46$_{-2.09}^{+2.32}$ & 0.52$_{-0.27}^{+0.53}$ & 10.20$_{-2.09}^{+2.26}$ & 3.13$_{-0.85}^{+0.86}$ & 3.58$_{-0.74}^{+0.80}$ & G & BW \\
h1 & PL & 5.67$_{-1.41}^{+1.62}$ & 2.94$_{-1.21}^{+1.74}$ & 9.16$_{-1.75}^{+2.48}$ & 0.71$_{-0.64}^{+0.72}$ & 3.23$_{-0.62}^{+0.87}$ & H & BW \\
h2 & APEC & 18.60$_{-2.46}^{+2.97}$ & 0.82$_{-0.23}^{+0.26}$ & 19.50$_{-2.44}^{+2.89}$ & 3.38$_{-0.33}^{+0.41}$ & 6.88$_{-0.86}^{+1.02}$ & -- & -- \\
j1 & NSA & 3.66$_{-1.46}^{+2.02}$ & 0.12$_{-0.10}^{+0.35}$ & 3.90$_{-1.52}^{+2.05}$ & 3.79$_{-1.73}^{+2.48}$ & 1.38$_{-0.54}^{+0.72}$ & J & I \\
k1 & PL & 18.60$_{-2.49}^{+2.96}$ & 19.00$_{-4.12}^{+5.05}$ & 40.90$_{-5.22}^{+6.54}$ & $-$0.02$_{-0.32}^{+0.32}$ & 14.40$_{-1.84}^{+2.31}$ & K & eBW \\
l1 & NSA & 10.70$_{-2.22}^{+2.92}$ & 0.76$_{-0.40}^{+0.72}$ & 11.60$_{-2.24}^{+2.84}$ & 2.87$_{-0.83}^{+0.90}$ & 4.09$_{-0.79}^{+1.00}$ & L & eBW \\
n1 & NSA & 3.89$_{-1.40}^{+2.19}$ & 0.34$_{-0.22}^{+0.45}$ & 4.37$_{-1.56}^{+2.22}$ & 2.56$_{-0.92}^{+1.36}$ & 1.54$_{-0.55}^{+0.78}$ & N & B \\
q1 & NSA & 4.17$_{-1.09}^{+1.23}$ & 0.68$_{-0.35}^{+0.45}$ & 4.86$_{-1.23}^{+1.49}$ & 1.90$_{-0.53}^{+0.83}$ & 1.72$_{-0.43}^{+0.53}$ & Q & B \\
r1 & NSA & 10.50$_{-4.27}^{+5.57}$ & 0.00$_{-0.00}^{+0.01}$ & 10.50$_{-4.28}^{+5.57}$ & 9.53$_{-1.91}^{+1.18}$ & 3.71$_{-1.51}^{+1.96}$ & -- & -- \\
r2 & NSA & 4.81$_{-2.56}^{+4.02}$ & 0.01$_{-0.01}^{+0.14}$ & 4.97$_{-2.62}^{+3.89}$ & 6.63$_{-3.30}^{+3.22}$ & 1.75$_{-0.92}^{+1.37}$ & -- & -- \\
\hline
\multicolumn{9}{p{12cm}}{{\it Notes}: $N_{\rm H}$ was fixed for all the fits at 1.05$\times$10$^{21}$ cm$^{-2}$. The errors are reported at 1-$\sigma$ confidence level.} \\
\multicolumn{9}{l}{$^a$ Preferred spectral model for each counterpart.} \\
\multicolumn{9}{p{12.5cm}}{$^b$ Unabsorbed fluxes in the bands 0.5--2 keV, 2--8 keV, and 0.5--10 keV, respectively, in units of 10$^{-16}$ erg~cm$^{-2}$~s$^{-1}$.} \\
\multicolumn{9}{l}{$^c$ X-ray colours of the counterparts, defined by $2.5\log{(F_{0.5-2}/F_{2-8})}$.} \\
\multicolumn{9}{p{12.5cm}}{$^d$ Unabsorbed luminosities in the band 0.5--10 keV, in  units of $10^{30}$ \ergs, assuming a distance of 5.43 kpc.} \\
\multicolumn{9}{p{12.5cm}}{$^e$ Confident X-ray counterparts to corresponding MSPs.} \\
\multicolumn{9}{p{12.5cm}}{$^f$ MSP types. I: isolated; B: binary; BW: non-eclipsing black widow; eBW: eclipsing black widow.} \\

    \end{tabular}
    \label{tab:X-ray_properties}
\end{table*}

\subsection{X-ray colour-magnitude diagram}
\label{subsec:cmd}

\begin{figure}
    \centering
    \includegraphics[width=\columnwidth]{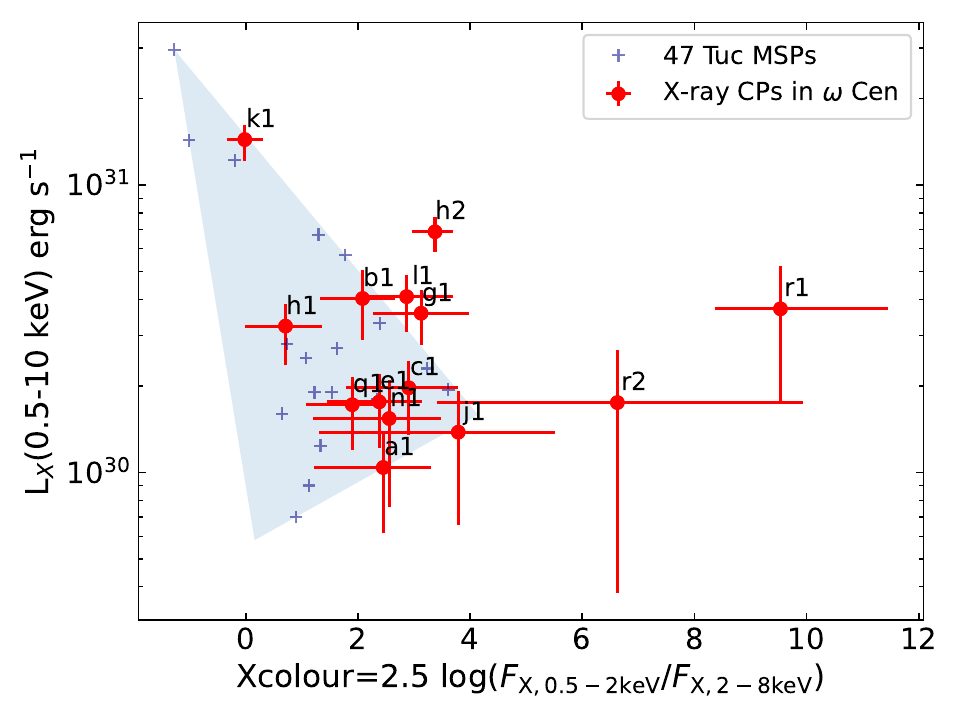}
    \caption{X-ray colour-magnitude diagram of 14 X-ray sources in \oc %in this work 
    (red dots with 1-$\sigma$ errorbars) and 18 MSPs in 47~Tuc (blue pluses). The shaded region indicates where MSPs %likely locate, 
    are generally observed, 
    based on the MSP population in 47~Tuc. The names of the X-ray sources in this work are %labelled correspondingly.
    indicated. 
    }
    \label{fig:cmd}
\end{figure}

To further %determine
verify 
the %real 
X-ray counterparts to MSPs in \oc, we %looked into the
created an 
X-ray colour-magnitude diagram (CMD) and compared the locations of %those 
our suggested 
X-ray counterparts with the locations of known X-ray MSPs in 47~Tuc in the CMD \citep[see][]{Lugger2023}. 
The X-ray colour is defined by Xcolour=$2.5\log{(F_{0.5-2}/F_{2-8})}$, where $F_{0.5-2}$ and $F_{2-8}$ are the unabsorbed fluxes in the bands 0.5--2 keV and 2--8 keV, respectively (see Table~\ref{tab:X-ray_properties}).
For the MSPs in 47~Tuc, we first extracted their counts from Table~5 in \citet{Heinke2005}, and then used the spectral fits in \citet{Bogdanov2006} and \citet{Bhattacharya2017} to convert counts into fluxes via {\sc pimms}\footnote{\url{https://cxc.harvard.edu/toolkit/pimms.jsp}}.
We finally obtained X-ray colours and luminosities for 18 MSPs in 47~Tuc. 
We show the locations of X-ray counterparts in this work (red dots) and MSPs in 47~Tuc (blue pluses) in the CMD in Figure~\ref{fig:cmd}, with a region indicating the likely distribution of MSPs.

Based on the CMD, it is clear that most X-ray counterparts are 
%likely 
plausible 
true matches to their corresponding MSPs, except for sources h2, r1, and r2, whose spectral fits are also not consistent with %typical 
the observed range of 
MSPs  \citep[see e.g.][]{Bogdanov2006}.
Therefore, besides the X-ray counterparts to MSPs A and B, which have been identified by \citet{Zhao2022}, we %newly identify confident 
suggest 
X-ray counterparts to nine MSPs in \oc (MSPs C, E, G, H, J, K, L, N, and Q).

\section{Discussion}
\label{sec:discussion}

\subsection{X-ray luminosities versus companion masses for spider MSPs}

It %is
has been 
found that redbacks (RBs) are relatively X-ray-bright with $L_X$ between $\sim2\times10^{31}$ \ergs and $3\times10^{32}$ \ergs, compared to black widows (BWs), with X-ray luminosities of $\sim2\times10^{30}$--$2\times10^{31}$ \ergs \citep{Roberts2014,Zhao2022}.
More intriguingly, a plausible bi-modal distribution of X-ray luminosities for eclipsing and non-eclipsing BWs was suggested, showing that eclipsing BWs are almost an order-of-magnitude %brighter 
more luminous 
than  non-eclipsing ones \citep{Zhao2022}. 
(Note that due to the absence of confirmed non-eclipsing RBs, the bimodality of RB X-ray luminosities is %plausible but is unable to be determined.)}
possible, but cannot be verified.)
If non-eclipsing BWs are essentially the same as eclipsing BWs, but seen at lower inclination, this implies that BW X-ray luminosity is substantially beamed in the equatorial plane. 
On the other hand, non-eclipsing BWs could be fundamentally lower in X-ray luminosity than eclipsing BWs. If they are observed at similar inclinations, then the measured mass function differences suggest that non-eclipsing BWs are lower mass than eclipsing BWs. This could explain the lower $L_X$ as due to a smaller wind mass loss from smaller (degenerate?) companions \citep[e.g.][]{Bailes11}.

In this work, we analysed the MSP population in \oc, which contains five BWs (MSPs B, K, and L are eclipsing BWs, and MSPs G and H are non-eclipsing; see Table~\ref{tab:X-ray_properties}), providing us a sample to %examine
test 
the bimodal distribution of X-ray luminosities for BWs. 
The eclipsing BW  
MSP K ($L_X=1.4_{-0.2}^{+0.2}\times10^{31}$ \ergs) is noticeably more luminous than the other four BWs, which have %X-ray luminosities of 
$L_X=3\times10^{30}$--$4\times10^{30}$ \ergs.
However, we cannot identify a statistically significant difference in X-ray luminosities between the eclipsing and non-eclipsing BWs in \oc\ (admittedly small numbers). 
Since the majority of X-ray emission from spider MSPs is considered to be produced by intrabinary shocks, which are generated from the collisions between the pulsar winds and mass outflow from the companion \citep[see e.g.,][]{Arons1993,Roberts2014,vanderMerwe2020}, brighter X-rays from spider MSPs indicate stronger intrabinary shocks produced in the binary systems.  
Moreover, \citet{Roberts2014} showed that RBs %generally 
are predicted to 
have more luminous shocks than BWs, which can be explained as more outflow material from more massive companions to collide with pulsar winds. 

\begin{figure*}
    \centering
    \includegraphics[height=0.8\columnwidth]{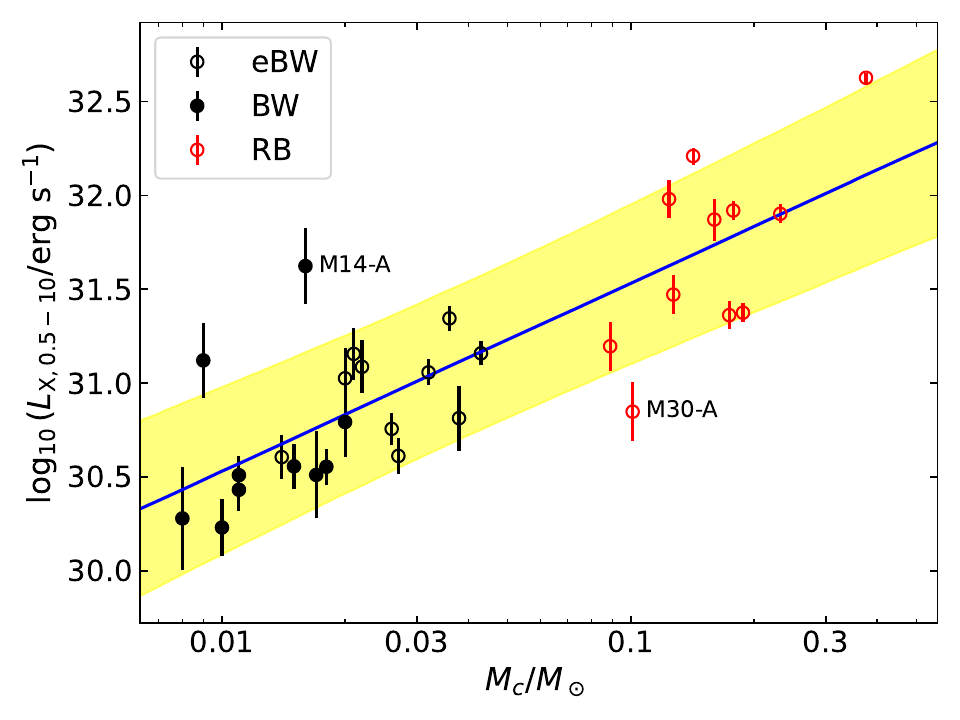}
    \includegraphics[height=0.8\columnwidth]{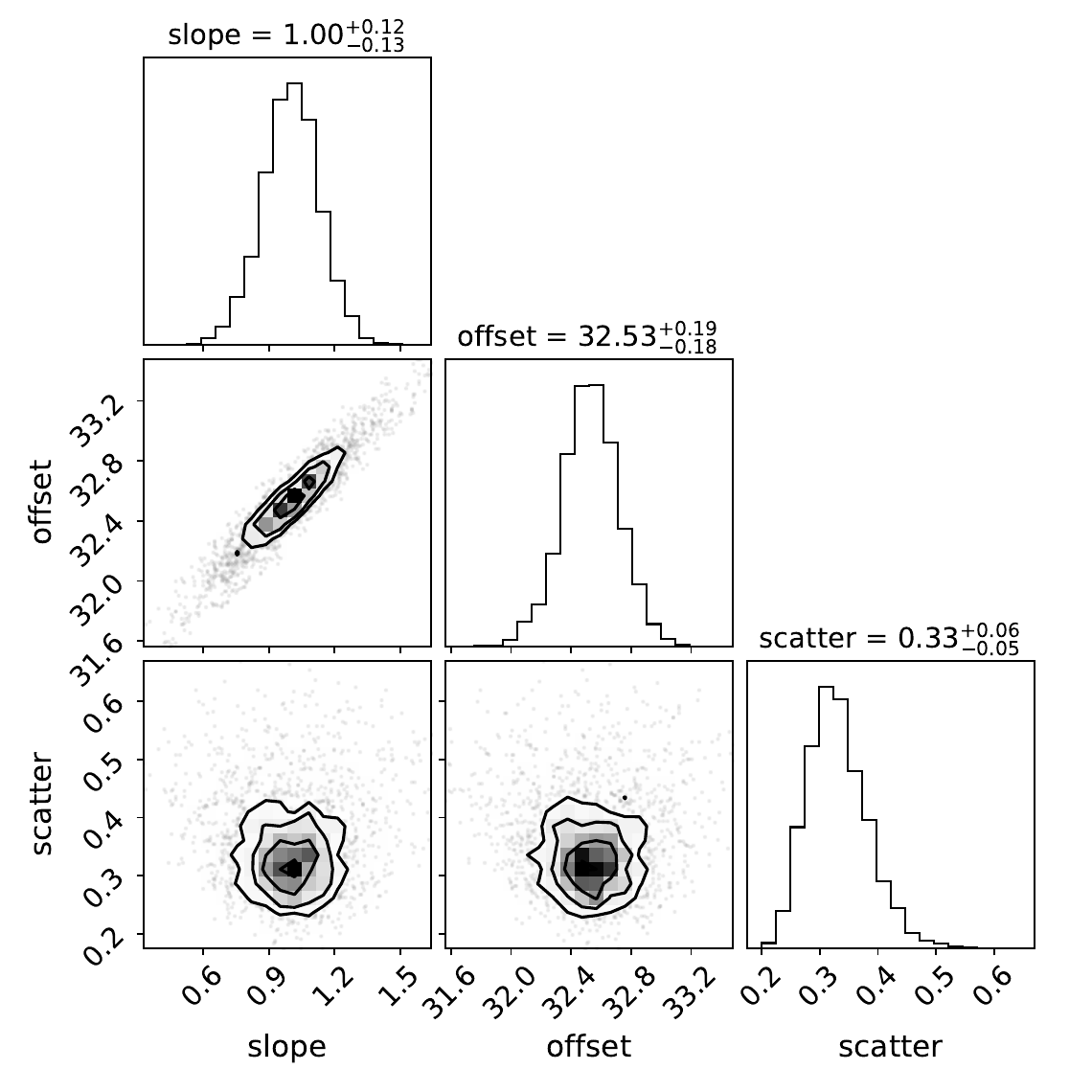}
    \caption{{\it Left}: X-ray luminosities of spider pulsars in the 0.5--10 keV band, versus the minimum masses of their companion stars, on a logarithmic scale. Red circles show redbacks, while black circles and filled dots show eclipsing and non-eclipsing black widows, respectively.
    Errorbars show 1-$\sigma$ uncertainties.
    The blue solid line indicates the best-fit function, whereas the yellow region shows the 1-$\sigma$ confidence %levels, 
    region, taking scatter into account. Two outliers, M14A and M30A, are labelled  (see text for more details). {\it Right}: The corner plot of the best-fit parameters, slope and offset, of the linear function shown in the left panel. Errors are presented at the 1-$\sigma$ confidence level. The correlation between X-ray luminosities and companion masses is clearly observed.}
    \label{fig:LxVSMc}
\end{figure*}

To investigate the influence of companion masses to X-ray luminosities of spider pulsars, we collected all the BWs and RBs in GCs with available companion masses and X-ray fluxes/luminosities published. 
(We also analysed the X-ray spectrum of M71E, which is a BW with a newly published timing solution \citep{Pan2023}. See Appendix~\ref{appendix:M71E}).
We note that it is very difficult to precisely measure the companion masses of spider pulsars in GCs (e.g. through optical observations of their light and radial velocity curves), given their low masses, faint magnitudes, and substantial distances. Among all the known spider pulsars in GCs, only a few have fairly accurate measurements of companion masses, such as M28H \citep{Pallanca2010}, M5C \citep{Pallanca2014}, and NGC6397A \citep{Ferraro2003}. For most spider pulsars, the companion masses are obtained by assuming inclination angles ($i$) and neutron star masses ($M_p$) of the binary systems, e.g., $i=90$\degree\ and $M_p=1.4$ \msun for minimum companion masses (see also below). However, we note that those better constrained companion masses are $\lesssim$30\% larger than the corresponding minimum companion masses. Hence, for the interest of this research, the companion masses reported here are all based upon the same assumptions to keep consistency.
Along with the results in this work, we compiled 11 RBs and 20 BWs with 10 eclipsing BWs and 10 non-eclipsing BWs (see Table~\ref{tab:spiders}). Note that all the observed RBs are eclipsing binaries. 
We plotted the X-ray luminosities in the band 0.5--10 keV versus the minimum companion masses of spider pulsars in Figure~\ref{fig:LxVSMc}. 
We found that a minimum mass of 0.019 M$_{\sun}$ divides most eclipsing BWs (above) from non-eclipsing BWs (below). 
The only exception in our catalogue is the eclipsing BW \oc-B, whose companion has a minimum mass of 0.013~\msun \citep[or a median mass of 0.016~\msun;][]{Dai2023}.
In addition, \citet{Dai2020} noted irregular eclipses observed in the orbital duration of \oc-B, which possibly suggests the inclination is low, such that we are seeing the ragged edge of the eclipse region.

It is clear that there is a correlation between X-ray luminosities and companion masses of spider pulsars (Pearson $r=0.81$, $p{\rm -value}=2.6\times10^{-8}$), such that spider pulsars with more massive companions are more luminous in X-rays (see Figure~\ref{fig:LxVSMc}). This is consistent with the evidence that RBs generally produce more energetic shocks than BWs \citep{Roberts2014}, and RBs are commonly an order-of-magnitude brighter than BWs \citep{Zhao2022}. 
Furthermore, we fitted a linear function through the data points (on a logarithmic scale) in Figure~\ref{fig:LxVSMc}, using the nested sampling algorithm UltraNest \citep{Buchner2021}. 
The best-fit function is given as $\log_{10}{L_X} = (1.0\pm0.1) \log_{10}{M_{c, min}} + (32.5\pm0.2)$, where $L_X$ is the X-ray luminosity in 0.5--10 keV, and $M_{c, min}$ is the minimum companion mass, 
plus an intrinsic scatter of $\log_{10}{L_X}$ of $\sim$0.3.
We also checked this correlation using X-ray luminosities in 0.3--8 keV, 
but the best fitted parameters remain consistent 
%with each other 
within the 1-$\sigma$ level. 

Almost all spider pulsars in our catalogue are consistent with the best-fit correlation within 
% 2-$\sigma$ 
the scatter region (see Figure~\ref{fig:LxVSMc}), with the exceptions of M14A, the brightest BW (non-eclipsing), and M30A, the faintest RB.
%The X-ray spectral analysis of M14A was presented by 
\citet{Zhao2022} showed that M14A's X-ray spectrum can be well-fitted by a single BB or NSA model, indicating principally thermal emission from M14A. 
The fitted PL model resulted in a photon index of 3.5$\pm$0.6 \citep{Zhao2022}, which also implies a significant thermal component in its emission. 
The lack of significant non-thermal emission from M14A might suggest either weak intrabinary shocks in this system, or a beaming effect of the shock emission. 
It also indicates the observed X-rays originate from the surface of the NS, i.e. the hotspots heated by returning relativistic particles from the magnetosphere. 
On the other hand, it is intriguing that M14A is such a bright, thermally-emitting MSP in X-rays, compared to other known thermally-emitting MSPs \citep[e.g.][]{Bogdanov2019}.
Another possibility is that M14A is a so-called non-eclipsing RB, rather than BW. In this case, its inclination angle is very low (close to face-on configuration), and therefore its true companion mass is much higher than the minimum companion mass, which is inferred by assuming an inclination of 90\degree. Supposing an inclination range between 10\degree\ and 5\degree, the corresponding companion mass of M14A is in the range of $\sim 0.1-0.2$ \msun, %belonging to RB population. 
which would %indicate it is
be consistent with 
a RB. 
However, it is important to note that the probability of inclination angles less than 10\degree\ is only about 1.5\%.
Thus, M14A might be a member of the missing non-eclipsing RBs, though the required inclination %is unusual.}
would be highly unlikely.
It is noteworthy, however, that there were only five photons available to perform spectral fitting of M14A \citep[see Figure~10 in][]{Zhao2022}, which %might produce biased fitting results. 
means the uncertainty in $L_X$ is substantial. 
Therefore, further X-ray observations of M14 are needed to better constrain and understand the X-ray properties of M14A and other MSPs in M14 \citep{Pan2021}. 

Another outlier is M30A, the X-ray faintest RB in our catalogue, with $L_X\sim7\times10^{30}$ \ergs in the band 0.5--10 keV \citep{ZhaoY2020}. 
In addition, its companion is the second lightest in our RB sample, with a minimum mass of $\sim$0.1 \msun \citep{Ransom2004}. 
However, Terzan 5A, the RB with the lightest companion \citep[$M_{c,min}\sim0.089$ \msun;][]{Ransom2005}, has a higher average X-ray luminosity of $\sim1.6\times10^{31}$ \ergs \citep{Bogdanov2021}, though highly variable (usually $L_X\sim10^{31}$ erg/s, except for 1-5 ks short flares reaching $L_X\sim 10^{33}$ erg/s). 
The large difference in X-ray luminosities of RBs with comparable companion masses implies other factor(s) may influence the observed X-ray emission, besides companion masses.
We suggest that the shock emission is likely beamed %in
along 
the shock surface, and hence the inclination to the shock surface also %matters
alters 
the received X-ray emission \citep[see e.g.,][]{Romani2016,Wadiasingh2017}. 

\begin{figure}
    \centering
    \includegraphics[width=\columnwidth]{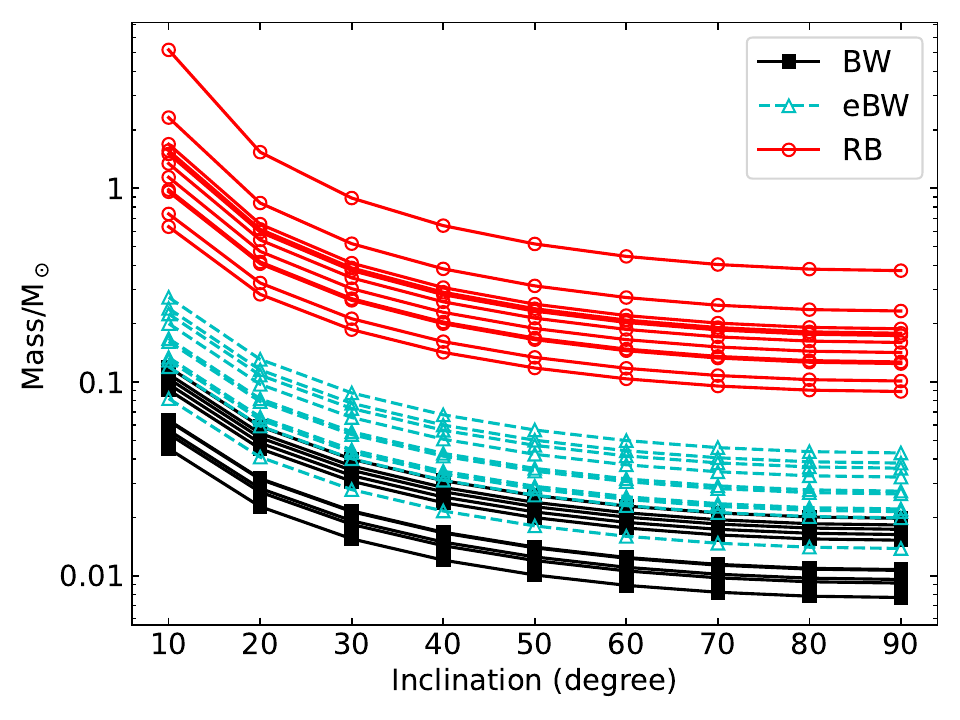}
    \caption{The relation between companion masses and inclination angles for given mass functions (see Eq.~\ref{eq:massfunc}) of the spider pulsars in our catalogue. Black squares and lines represent non-eclipsing black widows, cyan triangles and dashed lines represent eclipsing black widows, whereas red circles and lines show redbacks. The companion masses at an inclination of 90\degree\ show their minimum masses respectively. A clear gap between black widows and redbacks can also be seen from this figure.}
    \label{fig:mass-incl}
\end{figure}

It is also worth noting that the companion masses can be obtained from the binary mass function (in unit of solar mass)
\begin{equation}
    \label{eq:massfunc}
    f(M_c) = \frac{4\pi^2}{G} \frac{x^3}{{P_b}^2} = \frac{(M_c\sin{i})^3}{(M_c+M_p)^2},
\end{equation} 
where $x=a\sin{i}$ is the projected semi-major axis and $i$ is the inclination angle, $P_b$ is the orbital period, $M_c$ and $M_p$ are the companion mass and pulsar mass, respectively, and $G$ is the gravitational constant. 
The minimum companion masses are computed by assuming an inclination angle of 90\degree\ (i.e., edge-on) and a typical pulsar mass of 1.4~M$_{\sun}$. 
Given that inclination angles might be widely distributed, using minimum companion masses to explore the correlation with X-ray luminosities may introduce some bias.
For example, for a given mass function, the derived companion mass at an inclination of 45\degree\ is about 40\% to 60\% larger than its minimum companion mass. 
(However, this variation in implied mass due to different inclination angles would not be enough to explain the difference in companion masses between eclipsing and non-eclipsing black widows.)
In Figure~\ref{fig:mass-incl}, we illustrate the companion masses of our catalogued spider pulsars at different inclination angles based on their corresponding mass functions.

\begin{figure}
    \centering
    \includegraphics[width=\columnwidth]{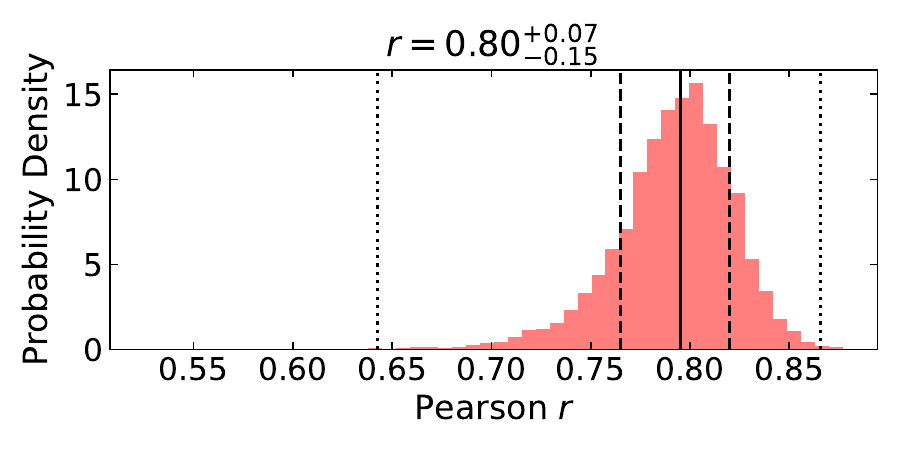}
    \includegraphics[width=\columnwidth]{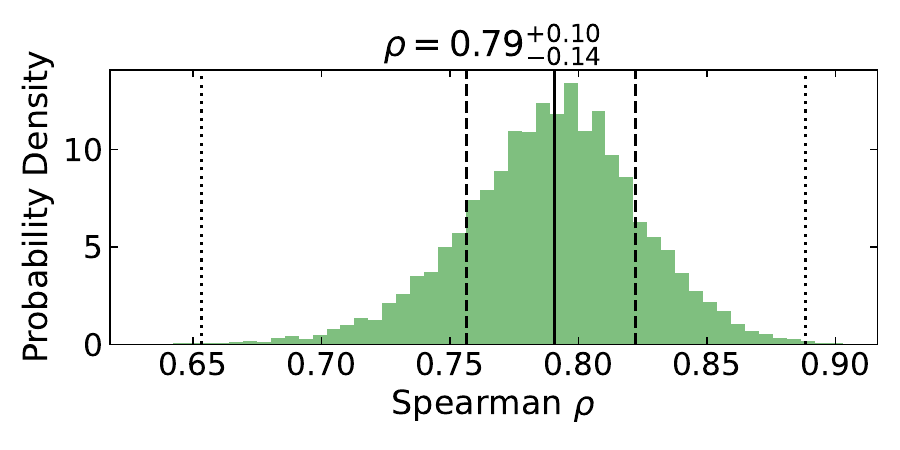}
    \includegraphics[width=\columnwidth]{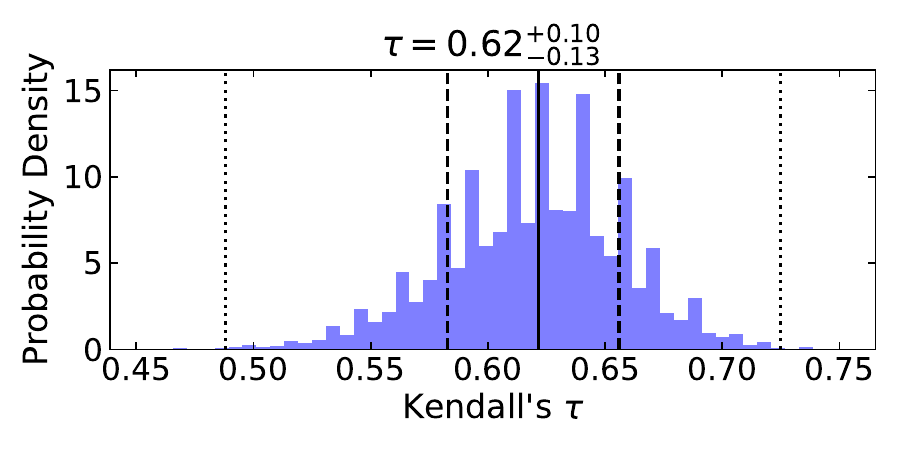}
    
    \caption{The distributions of Pearson $r$ (top panel), Spearman $\rho$ (middle panel), and Kendall's $\tau$ (bottom panel) correlation coefficients, respectively, which reflect the significance of correlation between X-ray luminosities and companion masses of spider pulsars in our work. These coefficients are obtained from 10,000 simulations of inclination angles for 31 spider pulsars in our work, which were used to derive companion masses. In each panel, the black solid, dashed, and dotted lines indicate the median value, 1-$\sigma$, and 3-$\sigma$ confidence levels, respectively, while the title shows the median value with 3-$\sigma$ confidence levels. It is clear that all of the three coefficients show strong evidence of a correlation of X-ray luminosities versus companion masses for spider pulsars.}
    \label{fig:coefficients}
\end{figure}

To thoroughly examine the correlation between X-ray luminosities and companion masses in our sample of spider pulsars, we implemented a rigorous approach involving simulations. We generated simulations by randomly selecting inclination angles for each spider pulsar based on a probability distribution derived from a sine function between 0 and $\pi$/2, as described by \citet{Backer1998}. Using the determined mass functions of spider pulsars (listed in Table~\ref{tab:spiders}), we calculated the corresponding companion masses using Equation~\ref{eq:massfunc}, assuming a fixed pulsar mass of 1.4 M$_{\sun}$.
Subsequently, utilising the simulated companion masses, we assessed the significance of the correlation between their simulated masses and $L_X$ by investigating the Pearson ($r$), Spearman ($\rho$), and Kendall's ($\tau$) correlation coefficients. We performed this calculation for a total of 10,000 iterations, generating distributions of these correlation coefficients (Figure~\ref{fig:coefficients}).
The obtained coefficients with 3-$\sigma$ confidence levels are $r=0.80^{+0.07}_{-0.15}$, $\rho=0.79^{+0.10}_{-0.14}$, and $\tau=0.62^{+0.10}_{-0.13}$, respectively, indicating a very strong positive correlation of $L_X$ versus $M_c$ (or a very strong linear correlation of $\log_{10}{L_X}$ versus $\log_{10}{M_c}$) for spider pulsars, even regardless of the inclination angles.
Therefore, our findings indicate that as the companion mass increases, the X-ray luminosity of the spider pulsar tends to increase as well.
It likely suggests that a more massive companion can produce stronger winds \citep[see e.g.][]{Bailes11}, and thus generate stronger intra-binary shocks with relativistic pulsar winds, leading to higher X-ray luminosities as observed. 

In addition, since the mass function contains information of both companion mass and inclination, which more generally reflects spider pulsar properties,
we therefore investigated the correlation between X-ray luminosities and mass functions of spider pulsars. 
We calculated their mass functions using Equation~\ref{eq:massfunc} with the values of $x$ and $P_b$ listed in Table~\ref{tab:spiders}. 
(It is noticeable that a recent work by \citealt{Koljonen2023} showed that there is no significant correlation between X-ray luminosities and orbital periods for spider pulsars.)
We found a significant correlation of X-ray luminosities with mass functions (both in logarithmic scale), with Pearson $r=0.81$, and $p{\rm -value}=2.8\times10^{-8}$. 
Using the same fitting prescription discussed above, we obtained a best-fit function of
$\log_{10}{L_X} = (0.35\pm0.04) \log_{10}{\rm MF} + (32.71\pm0.20)$,
where MF represents mass functions in  units of solar masses, with an intrinsic scatter of $\log_{10}{L_X}$ of $\sim$0.3. 
We suggest that both methods of using minimum companion masses and mass functions can provide reasonable predictions of X-ray luminosities for spider pulsars with radio timing solutions. 
However, the use of mass functions to infer X-ray luminosities may help mitigate some of the bias introduced by relying solely on minimum companion masses.

\defcitealias{Freire2017}{1}
\defcitealias{Heinke2005}{2}
\defcitealias{Bogdanov2006}{3}
\defcitealias{Edmonds2002}{4}
\defcitealias{Wang2020}{5}
\defcitealias{Zhao2021}{6}
\defcitealias{Pan2021}{7}
\defcitealias{Zhao2022}{8}
\defcitealias{Lynch2011}{9}
\defcitealias{Amato2019}{10}
\defcitealias{Begin2006}{11}
% Paulo Freire's GC Pulsar Catalogue (\url{https://www3.mpifr-bonn.mpg.de/staff/pfreire/GCpsr.html}){12}
\defcitealias{Vurgun2022}{13}
\defcitealias{Pallanca2010}{14}
\defcitealias{Pallanca2013}{15}
\defcitealias{Douglas2022}{16}
\defcitealias{Ransom2004}{17}
\defcitealias{ZhaoY2020}{18}
\defcitealias{Pallanca2014}{19}
% Zhang et al.(submitted){20}
\defcitealias{Lynch2012}{21}
\defcitealias{Oh2020}{22}
\defcitealias{Cocozza2008}{23}
\defcitealias{Cadelano2015}{24}
\defcitealias{Elsner2008}{25}
\defcitealias{Han2021}{26}
\defcitealias{Pan2023}{27}
% this work{28}
\defcitealias{Pan2020}{29}
\defcitealias{D'Amico2001}{30}
\defcitealias{Bogdanov2010}{31}
\defcitealias{Ferraro2001}{32}
\defcitealias{Ferraro2003}{33}
\defcitealias{Mucciarelli2013}{34}
\defcitealias{Zhang2022}{35}
\defcitealias{Pallanca2017}{36}
\defcitealias{Dai2023}{37}
\defcitealias{Chen2023}{38}
\defcitealias{Ransom2005}{39}
\defcitealias{Bogdanov2021}{40}
\defcitealias{Hessels2006}{41}

\begin{table*}
    \centering
    \caption{Properties of spider pulsars in GCs}
    \begin{tabular}{lrcccccccl}
\hline
MSP & Type$^a$ & ${P_b}^b$ & {\it x}$^c$ & ${\log_{10}{\rm MF}}^d$ & ${M_{c, min}}^e$ & ${M_{c, med}}^e$ & $\log_{10}L_{X, 0.3-8}$ & $\log_{10}L_{X, 0.5-10}$ & Reference\\
 & & (days) & (lt-s) & (\msun) & (\msun) & (\msun) & (\ergs) & (\ergs) & \\
\hline
47Tuc-J & eBW & 0.121 & 0.040 & $-5.313$ & 0.021 & 0.025 & $31.06 \pm 0.14$ & $31.16 \pm 0.14$ &  \citetalias{Freire2017,Heinke2005,Bogdanov2006}\\ %1,2,3  \\
47Tuc-O & eBW & 0.136 & 0.045 & $-5.272$ & 0.022 & 0.026 & $31.03 \pm 0.14$ & $31.09 \pm 0.14$ &  \citetalias{Freire2017,Heinke2005,Bogdanov2006}  \\
47Tuc-R & eBW & 0.066 & 0.033 & $-5.042$ & 0.026 & 0.031 & $30.85 \pm 0.09$ & $30.76 \pm 0.09$ &  \citetalias{Freire2017,Heinke2005,Bogdanov2006}  \\
47Tuc-W & RB & 0.133 & 0.243 & $-3.058$ & 0.127 & 0.148 & $31.42 \pm 0.10$ & $31.47 \pm 0.10$ & \citetalias{Freire2017,Heinke2005,Bogdanov2006,Edmonds2002}   \\
M13-E & eBW & 0.113 & 0.036 & $-5.409$ & 0.020 & 0.023 & $31.10 \pm 0.16$ & $31.03 \pm 0.16$ & \citetalias{Wang2020,Zhao2021} \\ %4,5   \\
M14-A & BW & 0.228 & 0.047 & $-5.666$ & 0.016 & 0.019 & $31.82 \pm 0.20$ & $31.62 \pm 0.20$ & \citetalias{Pan2021,Zhao2022} \\ %6,7 \\
M22-A & BW & 0.203 & 0.046 & $-5.585$ & 0.017 & 0.020 & $30.48 \pm 0.23$ & $30.51 \pm 0.23$ & \citetalias{Lynch2011,Amato2019} \\ %8,9  \\
M28-G & BW & 0.105 & 0.016 & $-6.357$ & 0.010 & 0.011 & $30.56 \pm 0.15$ & $30.23 \pm 0.15$ & \citetalias{Begin2006};~12;~\citetalias{Vurgun2022} \\ %10,11,12  \\
M28-H & RB & 0.435 & 0.719 & $-2.676$ & 0.174 & 0.203 & $31.24 \pm 0.20$ & $31.36 \pm 0.08$ & \citetalias{Begin2006};~12;~\citetalias{Vurgun2022,Pallanca2010} \\ %10,11,12,35  \\
M28-I* & RB & 0.459 & 0.766 & $-2.642$ & 0.178 & 0.209 & $32.34 \pm 0.08$ & $31.92 \pm 0.05$ & \citetalias{Begin2006};~12;~\citetalias{Vurgun2022,Pallanca2013} \\ %10,11,12,36  \\
M28-J & BW & 0.097 & 0.025 & $-5.752$ & 0.015 & 0.018 & $30.18 \pm 0.06$ & $30.56 \pm 0.12$ & \citetalias{Begin2006};~12;~\citetalias{Vurgun2022} \\ % 10,11,12  \\
M28-M & BW & 0.243 & 0.032 & $-6.205$ & 0.011 & 0.012 & $30.79 \pm 0.11$ & $30.43 \pm 0.11$ & \citetalias{Vurgun2022,Douglas2022} \\ %12,13  \\
M30-A & RB & 0.174 & 0.235 & $-3.338$ & 0.101 & 0.118 & $31.04 \pm 0.16$ & $30.85 \pm 0.16$ & \citetalias{Ransom2004,ZhaoY2020} \\ %14,15  \\
M5-C & eBW & 0.087 & 0.057 & $-4.572$ & 0.038 & 0.044 & $31.06 \pm 0.17$ & $30.81 \pm 0.17$ & \citetalias{Zhao2022,Pallanca2014};~20 \\ %7,16,17  \\
M5-G & BW & 0.114 & 0.036 & $-5.416$ & 0.020 & 0.023 & $30.87 \pm 0.19$ & $30.79 \pm 0.19$ &  20  \\
M62-B & RB & 0.145 & 0.253 & $-3.081$ & 0.124 & 0.145 & $32.00 \pm 0.10$ & $31.98 \pm 0.10$ & \citetalias{Lynch2012,Oh2020,Cocozza2008} \\ %18,19,37  \\
M71-A & eBW & 0.177 & 0.078 & $-4.785$ & 0.032 & 0.037 & $31.08 \pm 0.07$ & $31.06 \pm 0.07$ & \citetalias{Cadelano2015,Elsner2008} \\ %20,21  \\
M71-E & BW & 0.037 & 0.007 & $-6.635$ & 0.008 & 0.009 & $30.43 \pm 0.24$ & $30.28 \pm 0.27$ & \citetalias{Han2021,Pan2023};~24 \\ %22,23,24  \\
M92-A & RB & 0.201 & 0.399 & $-2.775$ & 0.160 & 0.187 & $31.92 \pm 0.11$ & $31.87 \pm 0.11$ & \citetalias{Zhao2022,Pan2020} \\ %7,25  \\
NGC6397-A & RB & 1.354 & 1.653 & $-2.578$ & 0.188 & 0.220 & $31.35 \pm 0.05$ & $31.37 \pm 0.05$ & \citetalias{D'Amico2001,Bogdanov2010,Ferraro2001,Ferraro2003,Mucciarelli2013} \\ %26,27,38,39,40  \\
NGC6397-B & RB & 1.977 & 2.580 & $-2.327$ & 0.232 & 0.273 & $31.83 \pm 0.05$ & $31.90 \pm 0.05$ & \citetalias{Bogdanov2010,Zhang2022,Pallanca2017} \\ %27,28,41  \\
NGC6544-A & BW & 0.071 & 0.012 & $-6.412$ & 0.009 & 0.011 & $31.00 \pm 0.20$ & $31.12 \pm 0.20$ & \citetalias{Zhao2022,Lynch2012} \\ %7,18  \\
\oc-B & eBW & 0.090 & 0.021 & $-5.879$ & 0.014 & 0.016 & $30.65 \pm 0.12$ & $30.61 \pm 0.12$ & \citetalias{Zhao2022};~24;~\citetalias{Dai2023} \\ %7,24,29  \\
\oc-G & BW & 0.109 & 0.032 & $-5.519$ & 0.018 & 0.021 & $30.65 \pm 0.10$ & $30.55 \pm 0.10$ & 24;~\citetalias{Chen2023} \\ %24,30  \\
\oc-H & BW & 0.136 & 0.022 & $-6.217$ & 0.011 & 0.012 & $30.63 \pm 0.10$ & $30.51 \pm 0.10$ &  24;~\citetalias{Chen2023} \\ %24,30  \\
\oc-K & eBW & 0.094 & 0.068 & $-4.418$ & 0.043 & 0.050 & $31.20 \pm 0.06$ & $31.16 \pm 0.06$ &  24;~\citetalias{Chen2023} \\ %24,30  \\
\oc-L & eBW & 0.159 & 0.062 & $-5.000$ & 0.027 & 0.032 & $30.69 \pm 0.10$ & $30.61 \pm 0.10$ & 24;~\citetalias{Chen2023} \\ %24,30   \\
Ter5-A$\dag$ & RB & 0.076 & 0.120 & $-3.493$ & 0.089 & 0.104 & $31.13 \pm 0.13$ & $31.20 \pm 0.13$ & \citetalias{Ransom2005,Bogdanov2021} \\ %31,32  \\
Ter5-O & eBW & 0.259 & 0.112 & $-4.650$ & 0.036 & 0.042 & $31.31 \pm 0.07$ & $31.34 \pm 0.07$ &  \citetalias{Ransom2005,Bogdanov2021} \\ %31,32  \\
Ter5-P & RB & 0.363 & 1.272 & $-1.775$ & 0.376 & 0.445 & $32.53 \pm 0.03$ & $32.63 \pm 0.03$ &  \citetalias{Ransom2005,Bogdanov2021} \\ %31,32  \\
Ter5-ad & RB & 1.094 & 1.103 & $-2.921$ & 0.142 & 0.165 & $32.14 \pm 0.05$ & $32.21 \pm 0.05$ &  \citetalias{Bogdanov2021,Hessels2006} \\ %32,33  \\
\hline
\multicolumn{10}{p{16cm}}{{\it Notes}: * M28-I %has been found switching 
switches 
between rotation-powered and accretion-powered states \citep{Papitto2013}. The X-ray luminosities reported here are based on the rotation-powered (or pulsar) state observed in 2015 \citep{Vurgun2022}. } \\
\multicolumn{10}{p{16cm}}{$\dag$ The X-ray luminosities of Terzan5-A reported here are time-averaged over low- and high-flux states. See \citet{Bogdanov2021} for more details. } \\
\multicolumn{10}{p{16cm}}{$^a$ %Types of spider MSPs. 
BW: non-eclipsing black widow; eBW: eclipsing black widow; RB: redback. Note that all RBs eclipse.} \\
\multicolumn{10}{p{16cm}}{$^b$ Orbital periods in units of days, obtained from Paulo Freire's GC Pulsar Catalogue (\url{https://www3.mpifr-bonn.mpg.de/staff/pfreire/GCpsr.html}).} \\
\multicolumn{10}{p{16cm}}{$^c$ Projected semimajor axis in units of light-seconds, obtained from Paulo Freire's GC Pulsar Catalogue.} \\
\multicolumn{10}{p{16cm}}{$^d$ Mass functions (MFs; in units of M$_{\sun}$) of spider pulsars in logarithm scale, calculated from $P_b$ and $x$ using Equation~\ref{eq:massfunc} } \\
\multicolumn{10}{p{16cm}}{$^e$ Minimum and median companion masses, assuming a neutron star mass of 1.4 M$_{\sun}$ and an inclination angle of 90\degree\ and 60\degree, respectively.} \\
\multicolumn{10}{p{16cm}}{References: 
(1) \citet{Freire2017}; 
(2) \citet{Heinke2005}; 
(3) \citet{Bogdanov2006}; 
(4) \citet{Edmonds2002};
(5) \citet{Wang2020}; 
(6) \citet{Zhao2021}; 
(7) \citet{Pan2021};
(8) \citet{Zhao2022};
(9) \citet{Lynch2011}; 
(10) \citet{Amato2019}; 
(11) \citet{Begin2006}; 
(12) Paulo Freire's GC Pulsar Catalogue (\url{https://www3.mpifr-bonn.mpg.de/staff/pfreire/GCpsr.html}); 
(13) \citet{Vurgun2022}; 
(14) \citet{Pallanca2010};
(15) \citet{Pallanca2013};
(16) \citet{Douglas2022}; 
(17) \citet{Ransom2004}; 
(18) \citet{ZhaoY2020};
(19) \citet{Pallanca2014}; 
(20) Zhang et al.(submitted); 
(21) \citet{Lynch2012}; 
(22) \citet{Oh2020}; 
(23) \citet{Cocozza2008};
(24) \citet{Cadelano2015}; 
(25) \citet{Elsner2008}; 
(26) \citet{Han2021}; 
(27) \citet{Pan2023}; 
(28) this work;
(29) \citet{Pan2020}; 
(30) \citet{D'Amico2001}; 
(31) \citet{Bogdanov2010}; 
(32) \citet{Ferraro2001};
(33) \citet{Ferraro2003};
(34) \citet{Mucciarelli2013};
(35) \citet{Zhang2022}; 
(36) \citet{Pallanca2017};
(37) \citet{Dai2023};  
(38) \citet{Chen2023}; 
(39) \citet{Ransom2005}; 
(40) \citet{Bogdanov2021}; 
(41) \citet{Hessels2006}.
} \\
    \end{tabular}
    \label{tab:spiders}
\end{table*}

\subsection{Applications}

An immediate application of the derived correlations between X-ray luminosities and minimum companion masses/mass functions is to predict X-ray luminosities of spider pulsars without determined X-ray counterparts yet, due to unsolved timing positions and/or limited X-ray %detectability. 
sensitivity. 
In Table~\ref{tab:predict_spider}, we collected %those
spider pulsars with reported binary parameters but without reported X-ray detections, and calculated their corresponding mass functions and companion masses. We then applied the correlation of $L_X$ versus mass function to predict their X-ray luminosity ranges (column 8 in Table~\ref{tab:predict_spider}).

Furthermore, based on the limiting X-ray luminosities (i.e. the lower limits of X-ray luminosities for sources to be detected) of corresponding GCs, we are able to predict if any of those spider pulsars might be detected using current X-ray observations.
Specifically, the limiting $L_X$ of 47~Tuc is about $3\times10^{29}$ \ergs (\citealt{Cheng2019}; see also Table 1 in \citealt{Zhao2022}), while the predicted $L_X$ of 47~Tuc I, P, V, ac, and ad are all well above this limit, and therefore their X-ray counterparts are expected to be detected in existing \chandra\ observations of 47~Tuc. In fact, all of these X-ray counterparts possibly have been tabulated in the existing \chandra\ surveys of 47 Tuc \citep[see e.g.][]{Heinke2005,Bhattacharya2017}, especially for 47 Tuc V and ad, whose X-ray luminosities are expected to be larger than $10^{31}$~\ergs.
%Particularly, 
47~Tuc I has a reported timing position, but is extremely close to another MSP, 47~Tuc G (0.12 arcsec apart in projected angular distance; \citealt{Freire2017}), impossible to spatially resolve with \chandra observations. %Alternatively, 
\citet{Bogdanov2006} identified and analysed one X-ray counterpart to 47~Tuc G and I %as a whole, 
together, 
yielding an unaborbed X-ray luminosity (0.3--8 keV) of $\sim 6 \times 10^{30}$ \ergs.

M3 A has no timing position published yet, while its X-ray counterpart is unlikely to be detected using currently available \chandra observations, since the upper limit of its predicted X-ray luminosity is  slightly lower than the limiting luminosity ($\sim 2\times10^{31}$ \ergs) of M3 \citep{ZhaoY2019}. This is  consistent with the results of X-ray source identifications in M3 conducted by \citet{ZhaoY2019}, who found none of the detected X-ray sources therein to be a confident MSP candidate. 

M62 E and F are of great interest, given their available timing solutions \citep{Lynch2012} and sufficient X-ray brightness (the limiting $L_X$ of M62 is $\sim3\times10^{30}$ \ergs; \citealt{Bahramian2020}). We noticed, however, that only M62 B and C have been identified with X-ray counterparts \citep[see][]{Oh2020}, whereas the other eight detected MSPs, four of which have published timing solutions, have no X-ray identifications. We briefly looked into the \chandra observations of M62, and found likely X-ray counterparts to M62 E and F, respectively. %, though their emission might be contaminated by neighbouring bright X-ray sources. 
Moreover, we searched for these two X-ray counterparts in the Goose Catalogue\footnote{\url{https://bersavosh.github.io/research/goose.html}} \citep{Bahramian2020}, where X-ray sources in M62 were catalogued. 
Based on the reported X-ray positions, we found two sources, CXOU J170113.24$-$300646.8 and J170112.77$-$300651.9 in their catalogue, likely match M62 E and F, respectively. 
The matching source to M62 E has an X-ray luminosity (0.5--10 keV) of $3.8^{+0.9}_{-0.8}\times10^{31}$ \ergs, fitted by a PL model (photon index $\Gamma=1.7_{-0.5}^{+0.6}$), whereas the upper limit of the predicted X-ray luminosity of M62 E is $\sim 2.8 \times 10^{31}$ \ergs, slightly lower than that of its matching source.
However, we note that M62 E is very close to a bright candidate CV (source 24 in \citealt{Oh2020}), and therefore, the X-ray emission from M62 E 
%is potentially 
could be 
contaminated by its neighbouring CV. Hence, CXOU J170113.24$-$300646.8 is still favoured as the genuine X-ray counterpart to M62 E.
On the other hand, CXOU J170112.77$-$300651.9, the matching source to M62 F, has an X-ray luminosity (0.5--10 keV) of $1.2^{+0.6}_{-0.4} \times 10^{31}$ \ergs, which falls into the predicted range of X-ray luminosity of M62 F. Plus, its fitted photon index is of $\Gamma=2.0^{+0.7}_{-0.5}$, typical for BWs. We thus suggest CXOU J170112.77$-$300651.9 is highly likely to be the X-ray counterpart to M62 F.

In addition, it is interesting to note that M62 C \citep{Possenti2003,Lynch2012} has an orbital period of $P_b\sim0.2$ days, similar to other known RBs, and a mass function of $\log_{10}{\rm MF}\sim -3.8$, which  is a factor of 2.7--57 lower than that of RBs with similar periods (such as M30 A and 47 Tuc V). 
No radio eclipses have been detected from M62 C \citep{Possenti2003}, but this might be explained by a low inclination (so our line of sight avoids the eclipsing material), which would also explain the low mass function. 
M62 C's X-ray luminosity (0.5--10 keV) of $5\times10^{31}$ \ergs\ \citep{Oh2020} is typical for RBs, supporting this identification. 
If M62 C has the same companion mass as M30 A, and M62 C simply has a lower inclination (assuming M30 A has a 60\degree\ inclination to our line of sight), we compute that M62 C would have an inclination of 37\degree. The probability of obtaining such a low inclination by chance is 20\%, which is quite plausible for a population of 17 known globular cluster redbacks. These inclinations and probabilities are probably overestimates, as M30 A has the lowest mass function among redbacks with similar orbital periods.
%Stringent
More detailed 
X-ray studies of MSPs in M62, and searches for radio eclipse signatures from M62 C, would be  productive future projects.

M14 D and E are two recently discovered redbacks with no timing positions %available 
yet \citep{Pan2021}. Although the limiting $L_X$ of M14 of $\sim 6 \times 10^{31}$ \ergs \citep{Bahramian2020} is relatively high, the predicted X-ray luminosities of M14 D and E suggest both MSPs may be detected with current \chandra observations. Indeed, Y. Zhao et al. (in preparation) used joint observations of \chandra, {\it HST}, and {\it VLA} of M14, and found one strong redback candidate and also another plausible redback candidate. %And it will be concluded with 
Future timing solutions of the MSPs in M14 can verify these candidates. 

Both NGC~6440 D (RB; \citealt{Freire2008}) and H (BW; \citealt{Vleeschower2022}) have reported timing positions, whereas their corresponding X-ray counterparts remain undetected. Due to occasional outbursts of X-ray transients in this cluster \citep[e.g.][]{in'tZand2001,Heinke2010}, only a few \chandra observations are available for faint X-ray source identifications. Due to obscuration and crowding in the core, the limiting $L_X$ is relatively high, at $\sim 4 \times 10^{31}$ \ergs\ \citep{Pooley2002}. Therefore, NGC~6440 H is unlikely to be detected given its X-ray luminosity is predicted to be much lower than the limiting $L_X$. Indeed, we checked the X-ray image of NGC~6440 and found no clear counterpart at the timing position of NGC~6440 H. We did not find a clear X-ray counter to NGC~6440 D either, likely implying its X-ray luminosity falls into the lower half of the prediction. 
It is also interesting to note that NGC~6440 D is the slowest rotating spider MSP in GCs found to date, with a spin period of 13.50~ms \citep{Freire2008}. The other two known `slow' spider MSPs, M30 A ($P\sim 11.02$~ms; \citealt{Ransom2004}) and Terzan~5 A ($P\sim 11.56$~ms; \citealt{Ransom2005}), are the faintest RBs in X-rays in our catalogue. The X-ray-faint nature of these slow rotating spider MSPs might stem from weak 
pulsar winds produced by the neutron star, which as a result weakens intra-binary shocks. The X-ray luminosity of NGC~6440 D, therefore, might be even lower than our prediction. However, more samples of slow rotating spider MSPs would be needed to confirm this trend.

M28 L and N are located in the crowded core region of M28 (see e.g. Figure~1 in \citealt{Vurgun2022}), adjacent to bright X-ray sources. Thus, even though these two timing-solved BWs are predicted to be brighter than the limiting $L_X$ of M28 of $\sim 8 \times 10^{29}$ \ergs \citep{Bahramian2020}, their emission is highly likely to be contaminated and/or concealed by their neighbouring bright sources. For example, \citet{Bogdanov2011} analysed the potential X-ray counterpart to M28 L and suggested the X-rays extracted around M28 L were most likely dominated by unrelated source(s) due to source blending.
More recently,
\citet{Vurgun2022} claimed an identification of the X-ray counterpart to M28 L, %\citep[see also][]{Bogdanov2011}, 
and %found an X-ray luminosity (0.5--10 keV) of it to be $\sim 2 \times 10^{32}$ \ergs. 
measured its $L_X$(0.5--10 keV)$\sim 2 \times 10^{32}$ \ergs. 
However, it would be very unusual if this X-ray source is the true X-ray counterpart to M28 L, since non-eclipsing BWs typically have X-ray luminosities lower than $1\times10^{31}$ \ergs (see Figure~\ref{fig:LxVSMc}). 
Therefore, the X-ray counterpart to M28 L identified by \citet{Vurgun2022} is probably not genuine. And we suggest that M28 L and N are hardly detectable with current \chandra observations.

The X-ray counterparts to NGC~6624 F and NGC~6712 A are not detectable, since their emission is completely screened by persistent luminous LMXBs (peak $L_X \gtrsim 10^{35}$ \ergs) in their corresponding clusters (see Table 7 in \citealt{Bahramian2023}). 
Finally, the limiting $L_X$ of NGC~6760 is about $1\times10^{31}$ \ergs \citep{Bahramian2020}, and thus NGC~6760 A is more likely undetectable based on its predicted X-ray luminosity. Since NGC~6760 A has a published timing position \citep{Freire2005}, we simply looked into the \chandra observation of NGC~6760, and found no clear counterpart at its timing position, which is consistent with our prediction.

\begin{table*}
    \centering
    \caption{Spider pulsars with predicted X-ray luminosities in GCs}
    \begin{tabular}{lrccccccl}
\hline
MSP & Type & ${P_b}$ & {\it x} & ${\log_{10}{\rm MF}}$ & ${M_{c, min}}$ & ${M_{c, med}}$ & $\log_{10}L_{X, 0.5-10}$* & Reference\\
 & & (days) & (lt-s) & (\msun) & (\msun) & (\msun) & (\ergs) & \\
\hline
47Tuc-I	&	BW	&	0.230	&	0.038	&	$-$5.938	&	0.013	&	0.015	&	[30.21, 31.08]	& 1	\\
47Tuc-P	&	BW	&	0.147	&	0.038	&	$-$5.566	&	0.018	&	0.020	&	[30.35, 31.20]	& 2	\\
47Tuc-V	&	RB	&	0.212	&	0.742	&	$-$2.012	&	0.305	&	0.359	&	[31.53, 32.45]	& 2	\\
47Tuc-ac	&	eBW	&	0.179	&	0.020	&	$-$6.607	&	0.008	&	0.009	&	[29.96, 30.87]	& 3	\\
47Tuc-ad	&	RB	&	0.318	&	0.682	&	$-$2.475	&	0.205	&	0.240	&	[31.38, 32.28]	& 3	\\
M3-A	&	BW	&	0.136	&	0.026	&	$-$6.011	&	0.012	&	0.014	&	[30.18, 31.06]	& 4	\\
M62-E	&	eBW	&	0.158	&	0.070	&	$-$4.831	&	0.031	&	0.036	&	[30.61, 31.44]	& 5	\\
M62-F	&	BW	&	0.205	&	0.057	&	$-$5.321	&	0.021	&	0.025	&	[30.43, 31.28]	& 5	\\
M14-D	&	RB	&	0.743	&	0.755	&	$-$3.078	&	0.125	&	0.145	&	[31.19, 32.06]	& 4	\\
M14-E	&	RB	&	0.846	&	1.102	&	$-$2.698	&	0.170	&	0.199	&	[31.31, 32.20]	& 4	\\
NGC6440-D	&	RB	&	0.286	&	0.397	&	$-$3.086	&	0.124	&	0.144	&	[31.19, 32.05]	& 6	\\
NGC6440-H	&	BW	&	0.361	&	0.025	&	$-$6.874	&	0.006	&	0.007	&	[29.86, 30.79]	& 7	\\
NGC6624-F	&	RB	&	0.221	&	0.285	&	$-$3.291	&	0.105	&	0.122	&	[31.12, 31.98]	& 3	\\
M28-L   &   BW      &   0.226    &   0.057   &   $-$5.409    &    0.020   &   0.023   &   [30.40, 31.25]  &   8,9 \\
M28-N	&	BW	&	0.198	&	0.050	&	$-$5.475	&	0.019	&	0.022	&	[30.38, 31.23]	& 10	\\
NGC6712-A	&	BW	&	0.148	&	0.049	&	$-$5.236	&	0.023	&	0.026	&	[30.46, 31.31]	& 11	\\
NGC6760-A	&	BW	&	0.141	&	0.038	&	$-$5.541	&	0.018	&	0.021	&	[30.36, 31.21]	& 12	\\
\hline
\multicolumn{9}{p{12.5cm}}{* Unabsorbed X-ray luminosities predicted using the correlation between $L_X$ and MF (see text for details). The presented ranges of X-ray luminosities take both intrinsic scatter and 1-$\sigma$ uncertainty into account.} \\
\multicolumn{9}{p{12.5cm}}{
Reference:
(1) \citet{Freire2017};
(2) \citet{Ridolfi2016};
(3) \citet{Ridolfi2021};
(4) \citet{Pan2021};
(5) \citet{Lynch2012};
(6) \citet{Freire2008};
(7) \citet{Vleeschower2022};
(8) Paulo Freire's GC Pulsar Catalogue (\url{https://www3.mpifr-bonn.mpg.de/staff/pfreire/GCpsr.html});
(9) \citet{Bogdanov2011};
(10) \citet{Douglas2022};
(11) \citet{Yan2021};
(12) \citet{Freire2005}.
} \\
\end{tabular}
\label{tab:predict_spider}
\end{table*}

\section{Conclusions}
\label{sec:conclusions}

In this work, we presented a comprehensive X-ray study of 
the known 
MSPs in the globular cluster \oc. 
Utilising the latest published radio detections and timing solutions of the 18 MSPs in \oc \citep{Chen2023,Dai2023}, we conducted a targeted search for their X-ray counterparts using four \chandra observations with a total exposure time of 290.9 ks. 
Our analysis yielded 14 X-ray sources in the vicinities of the radio positions of 12 MSPs, with seven newly detected sources and seven sources previously detected \citep{Haggard2009,Henleywillis2018}.

By performing X-ray spectral fitting analyses for these sources, we found that most of them %X-ray sources 
can be well described by a neutron star hydrogen atmosphere model \citep{Heinke2006}, with exceptions such as sources h1, k1, and h2, which are best fitted by a power-law model and a %diffuse gas 
thermal plasma 
emission model, respectively. 
Furthermore, taking into account the X-ray colours of the sources and their positions in the colour-magnitude diagram, we identified 11 confident X-ray counterparts to the MSPs in \oc, of which nine (MSPs \oc C, E, G, H, J, K, L, N, and Q) were newly identified in this work. 
The X-ray luminosities of these identified MSPs range from $1.0\times10^{30}$ \ergs to $1.4\times10^{31}$ \ergs, in the band 0.5--10 keV. 
Notably, MSP \oc K has an X-ray luminosity of $\sim1.44\times10^{31}$ \ergs, making it one of the most X-ray-luminous black widow MSPs in Galactic globular clusters.

%In addition, we studied
We also studied (Appendix A)
 the X-ray counterpart to MSP E (or M71E) in the GC M71, which has the shortest orbital period among all discovered MSP binaries found so far \citep{Han2021,Pan2023}.
Using the 52.5-ks \chandra\ observation of M71, we extracted and analysed the X-ray spectrum of M71E. The X-ray spectrum of M71E is well described by blackbody-like models (BB or NSA), and we determined its unabsorbed X-ray luminosity to be $1.9^{+1.2}_{-1.1} \times 10^{30}$ \ergs within the 0.5--10 keV range, at a distance of 4 kpc.

We %also
conducted an empirical investigation of the correlation between X-ray luminosities and companion masses as well as mass functions
of spider pulsars in globular clusters. 
Our analysis yielded a best-fit function of 
$\log_{10}{L_X} = (1.0\pm0.1) \log_{10}{M_{c, min}} + (32.5\pm0.2)$, where $L_X$ is the X-ray luminosity in 0.5--10 keV, and $M_{c, min}$ is the minimum companion mass. 
For the mass functions, we found a relation of 
$\log_{10}{L_X} = (0.35\pm0.04) \log_{10}{\rm MF} + (32.71\pm0.20)$, where MF denotes mass functions in the units of solar mass. 
Both fittings yield an intrinsic scatter of $\log_{10}{L_X}$ of $\sim$0.3.
%Additionally, 
We suggested that the inclination to the intrabinary shocks may also influence the observed X-ray emission from spider MSPs, in addition to the companion masses.

\section*{Acknowledgements}

We thank the referee for exceptionally useful and constructive comments.
We thank Weiwei Chen for useful discussions of the radio localisation of MSP H.
CH is supported by NSERC Discovery Grant RGPIN-2016-04602. 
JZ is supported by China Scholarship Council (CSC), File No. 202108180023. 
This research has made use of data obtained from the Chandra Data Archive, and software provided by the Chandra X-ray Centre (CXC) in the application packages {\sc ciao}, {\sc sherpa}, {\sc ds9}, and {\sc pimms}.
This work has used TOPCAT, an interactive graphical viewer and editor for tabular data \citep{Taylor2005}. 
This research has made use of the VizieR catalogue access tool, CDS,
 Strasbourg, France (DOI: 10.26093/cds/vizier). The original description 
 of the VizieR service was published by \citet{Ochsenbein2000}. %2000, A\&AS 143, 23.
This research has made use of NASA’s Astrophysics Data System.

%%%%%%%%%%%%%%%%%%%%%%%%%%%%%%%%%%%%%%%%%%%%%%%%%%
\section*{Data Availability}

The X-ray data underlying this article are available in the Chandra Data Archive (\url{https://cxc.harvard.edu/cda/}) by searching the Observation IDs listed in Table~\ref{tab:obs_oc} in the Search and Retrieval interface, ChaSeR (\url{https://cda.harvard.edu/chaser/}).



\bibliographystyle{mnras}
\bibliography{ref} % if your bibtex file is called example.bib

%\begin{thebibliography}{99}
%\end{thebibliography}

%%%%%%%%%%%%%%%%%%%%%%%%%%%%%%%%%%%%%%%%%%%%%%%%%%

%%%%%%%%%%%%%%%%% APPENDICES %%%%%%%%%%%%%%%%%%%%%

\appendix

\section{X-ray spectral analysis of PSR M71E}
\label{appendix:M71E}

PSR M71E (or PSR J1953+1846E) is a non-eclipsing BW in the globular cluster M71 (NGC 6838) recently discovered in the FAST Galactic Plane Pulsar Snapshot survey \citep{Han2021}.
The latest published timing solution for this system unveils that it has the shortest orbital period among all known MSP binaries to date \citep[$P_b \sim$ 53 minutes;][]{Pan2023}.
The X-ray counterpart to M71E has been catalogued by \citet{Elsner2008} in their Table 1 (Source s34), with only a few X-ray properties presented. 
On the other hand, \citet{Pan2023} roughly estimated the X-ray luminosity of M71E to be $6 \pm 2 \times 10^{29}$ \ergs.
However, its X-ray luminosity was likely underestimated, given that the limiting unabsorbed X-ray luminosity of M71 is about $2 \times 10^{30}$ \ergs \citep[0.3--8 keV;][]{Zhao2022}.
Therefore, to better constrain its X-ray properties and to enhance our sample of spider pulsars, we conducted a comprehensive X-ray analysis of M71E in this appendix.

We obtained the %only
\chandra\ observation of M71 from the \chandra\ Data Archive\footnote{\url{https://cxc.cfa.harvard.edu/cda/}} (Obs. ID 5434 with an exposure time of 52.45~ks), and reprocessed the dataset using the {\tt chandra\_repro} script.
We extracted the X-ray spectrum of M71E within a 1-arcsec-radius circle, centred at its timing position \citep[RA=19:53:37.9464, Dec.=+18:44:54.310;][]{Pan2023}, while the background spectrum was extracted from a source-free region around M71E.

We fitted the X-ray spectrum of M71E with an absorbed BB, NSA, and PL, respectively. The $N_{\rm H}$ towards M71 was estimated and fixed to be 2.18$\times$10$^{21}$ cm$^{-2}$ (see Section~\ref{subsec:spectra} for the estimation method), and the distance to M71 was assumed to be 4.0 kpc \citep[2010 edition]{Harris1996}. We summarised the fitting results in Table~\ref{tab:spec_fits_M71E}. 
%It is found that 
The X-ray spectrum of M71E can be statistically well fitted by all the three models. 
However, like the spectra of BWs in \oc, the relatively high photon index ($\Gamma= 2.9^{+0.8}_{-0.7}$) %fitted 
inferred 
from the PL model implies thermal X-ray emission is %more significant from 
a more physically appropriate model for 
M71E. Hence, blackbody-like spectral models (BB or  NSA) are preferred for M71E.
By adopting the NSA model fits, the unabsorbed X-ray luminosity of M71E is found to be $1.9^{+1.2}_{-1.1} \times 10^{30}$ \ergs in 0.5--10 keV (or $2.7^{+1.5}_{-1.5} \times 10^{30}$ \ergs in 0.3--8 keV). 

\begin{table}
    \caption{Spectral fits for PSR J1953+1846E in M71.}
    \centering
    \begin{tabular}{lccc}
      \hline
       & & Spectral Model & \\
       & BB & NSA & PL  \\
      \hline
      {$kT_{\rm BB}$/$\log T_{\rm eff}$}/$\Gamma$\,$^a$ & $0.19^{+0.05}_{-0.04}$ & $6.07_{-0.18}^{+0.20}$ & $2.9^{+0.8}_{-0.7}$ \\
      Reduced Stat. & 0.91 & 0.96 & 1.03 \\
      Q-value & 0.51 & 0.46 & 0.41 \\
      {$F_X$(0.5--10 keV)}$^b$ & $1.1^{+0.6}_{-0.5} $ & $1.0^{+0.6}_{-0.6}$ & $1.9^{+0.6}_{-0.6}$ \\
      \hline
    \multicolumn{4}{p{0.9\linewidth}}{{\it Notes}: $N_{\rm H}$ was fixed for all fits at $2.18 \times 10^{21}~$cm$^{-2}$.} \\
    \multicolumn{4}{p{0.9\linewidth}}{$^a$ $kT_{\rm BB}$: blackbody temperature in units of keV; $\log T_{\rm eff}$: unredshifted effective temperature of the NS surface in units of log Kelvin; $\Gamma$: photon index.} \\
    \multicolumn{4}{l}{$^b$ Unabsorbed flux in units of $10^{-15}$ erg cm$^{-2}$~s$^{-1}$.} \\
    \end{tabular}
    \label{tab:spec_fits_M71E}
\end{table}

\begin{figure}
    \centering
    \includegraphics[width=\columnwidth]{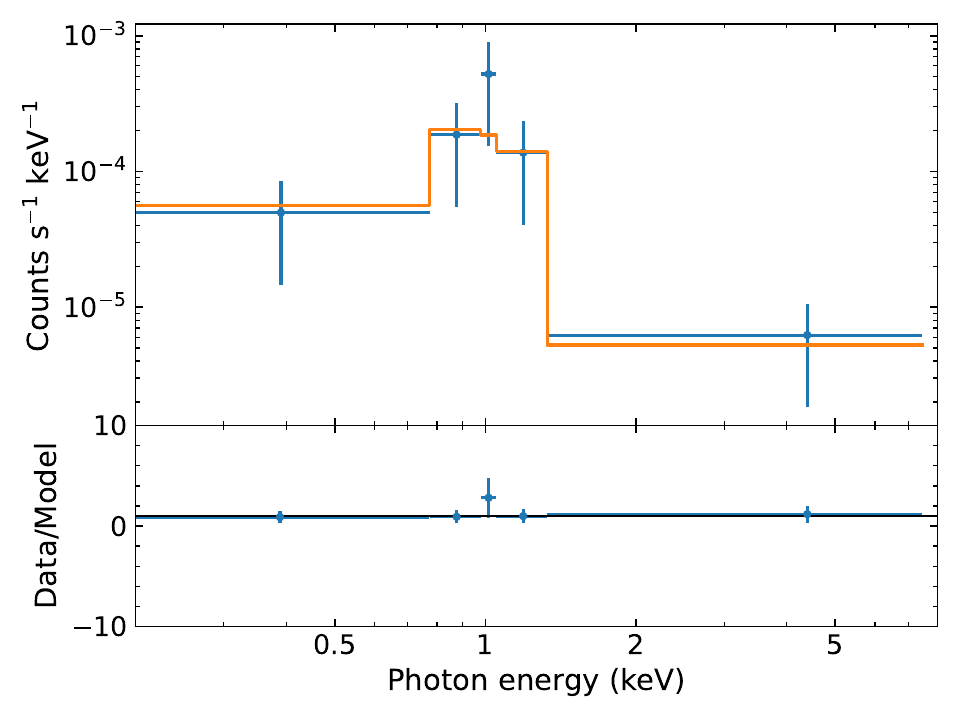}
    \caption{X-ray spectrum and the best-fitted NSA model of M71E in the energy range 0.3--8 keV. The data are grouped to a signal-to-noise ratio of at least 1 per bin for display purposes, while data were grouped to at least 1 photon per bin for fitting processes. The spectrum was fitted with WSTAT statistics. }
    \label{fig:M71E}
\end{figure}

% \section{Some extra material}

% If you want to present additional material which would interrupt the flow of the main paper,
% it can be placed in an Appendix which appears after the list of references.

%%%%%%%%%%%%%%%%%%%%%%%%%%%%%%%%%%%%%%%%%%%%%%%%%%

% Don't change these lines
\bsp	% typesetting comment
\label{lastpage}
\end{document}